\documentclass[acmsmall,screen]{acmart}
\acmJournal{TOSEM}

\usepackage[utf8]{inputenc}
\usepackage[english]{babel}

\usepackage{graphicx}
\usepackage{placeins}
\DeclareGraphicsExtensions{.pdf,.png,.jpg}

\usepackage{booktabs}
\usepackage{array}
\usepackage{enumitem}
\usepackage{makecell}
\usepackage[ruled,vlined,linesnumbered]{algorithm2e}

\usepackage{xcolor}
\usepackage[most]{tcolorbox}
\usetikzlibrary{arrows.meta}

\definecolor{darkgreen}{RGB}{0,100,0}
\definecolor{codegreen}{rgb}{0,0.5,0}
\definecolor{codepurple}{rgb}{0.58,0,0.82}
\definecolor{codegray}{rgb}{0.5,0.5,0.5}
\usepackage{listings}
\lstdefinestyle{mystyle}{
  commentstyle=\color{codegreen},
  keywordstyle=\bfseries,
  stringstyle=\color{codepurple},
  basicstyle=\ttfamily\scriptsize,
  breaklines=true,
  captionpos=b,
  keepspaces=true,
  tabsize=2
}
\lstdefinestyle{rqcode}{
  style=mystyle,
  backgroundcolor=\color{black!4!white},
  commentstyle=\itshape\color{black!58!white},
  frame=none,
  aboveskip=4pt,
  belowskip=4pt,
  xleftmargin=6pt,
  xrightmargin=6pt,
  showstringspaces=false,
  columns=fullflexible
}
\lstdefinestyle{prompt}{
  style=mystyle,
  basicstyle=\ttfamily\fontsize{7.1}{8.5}\selectfont,
  backgroundcolor=\color{black!4!white},
  frame=leftline,
  rulecolor=\color{black!35!white},
  framerule=0.7pt,
  framesep=5pt,
  xleftmargin=7pt,
  xrightmargin=3pt,
  aboveskip=3pt,
  belowskip=7pt,
  breaklines=true,
  breakatwhitespace=false,
  columns=fullflexible,
  keepspaces=true,
  showstringspaces=false
}
\newcommand{\find}[1]{%
\begin{tcolorbox}[tile,size=fbox,boxsep=2mm,boxrule=0pt,top=0pt,bottom=0pt,
borderline={0.6mm}{0pt}{black!66!white},colback=black!5!white]
\em #1
\end{tcolorbox}
}

\newtcolorbox{rqtwopathbox}[1][]{
  colback=black!2!white,colframe=black!50!white,boxrule=0.4pt,
  arc=0pt,left=4pt,right=4pt,top=3pt,bottom=3pt,
  before skip=4pt,after skip=4pt,breakable,#1
}
\newcommand{\rqtwocritical}[1]{%
  \textbf{#1}%
}

\newenvironment{rqcodeblock}[1][0.88\linewidth]{%
  \begin{center}\begin{minipage}{#1}%
}{%
  \end{minipage}\end{center}%
}

\definecolor{DarkOrange}{rgb}{0.8,0.3,0.0} 
\definecolor{DarkCyan}{rgb}{0.0, 0.55, 0.55}
\definecolor{codegreen}{rgb}{0,0.6,0}
\definecolor{codegray}{rgb}{0.5,0.5,0.5}
\definecolor{codepurple}{rgb}{0.58,0,0.82}
\definecolor{backcolour}{rgb}{0.95,0.95,0.92}

\newcommand{\papertitle}{\tech: Characterizing Tensor Binding Failures in AI Compiler and Runtime Stacks}
\newcommand{\papertitleshort}{\tech: Characterizing Tensor Binding Failures}

\newcommand{\tech}{\mbox{\textsc{TensorBound}}}

\newcommand{\paperkeywords}{AI compilers, accelerator runtimes, tensor storage reuse, empirical software engineering, program analysis}

\newcommand{\authorAname}{Baihong Chen}
\newcommand{\authorAaffil}{Utah State University}
\newcommand{\authorAemail}{b.chen@usu.edu}

\newcommand{\authorBname}{Yongxi Sha}
\newcommand{\authorBaffil}{Utah State University}
\newcommand{\authorBemail}{yongxi.sha@usu.edu}

\newcommand{\authorCname}{Wen Li}
\newcommand{\authorCaffil}{Utah State University}
\newcommand{\authorCemail}{awen.li@usu.edu}

\newcommand{\appendixcontent}{%
  \appendix
}

\newcommand{\artifact}[1]{%
  \href{https://github.com/awen-li/TensorBound/tree/b1ba851239c343799d8678f29465fe5f90ec4b1c}{#1}%
}

\setcopyright{acmcopyright}

\begin{document}

\title[\papertitleshort]{\papertitle}

\author{\authorCname}
\affiliation{%
  \institution{\authorCaffil}
  \city{Logan}
  \state{Utah}
  \country{USA}
}
\email{\authorCemail}

\author{\authorBname}
\affiliation{%
  \institution{\authorBaffil}
  \city{Logan}
  \state{Utah}
  \country{USA}
}
\email{\authorBemail}

\author{\authorAname}
\affiliation{%
  \institution{\authorAaffil}
  \city{Logan}
  \state{Utah}
  \country{USA}
}
\email{\authorAemail}



\begin{abstract}
AI compiler and runtime stacks must preserve bindings between
tensor values and their operations, storage, and execution state.
These bindings can fail while the underlying objects remain
present, with symptoms appearing in another layer.
We present \tech{}, an empirical study of tensor binding failures
in IREE and the XLA and PyTorch ecosystems.
The study analyzes 264 reports and reconstructs 149 distinct
repaired causes through source tracing and comparisons of
pre-fix and repaired revisions.
Each reconstruction connects the first invalid decision to
its violated condition, physical effect, and repair.
Missing eligibility checks, incorrect mappings or translations,
and lost or omitted relations account for {69.1\%} of repaired causes.
These patterns highlight the need to preserve binding facts
and check the conditions for their use.
The 101 first invalid decisions in compiler code identify
candidate checking points before execution.
Four archived IREE diagnostic attempts distinguish effect
localization from source-based cause explanation.
Together, these findings motivate transition-specific checks
that preserve binding facts across representations and enforce
the required conditions at consuming decisions.
\end{abstract}


\maketitle

\section{Introduction}\label{sec:intro}

AI compiler and runtime stacks translate logical tensor programs
into operations on physical storage.
Graph and tensor IR represent operations and their data dependencies.
Compiler transformations restructure these operations, while
bufferization and memory planning map tensor values to storage regions.
These decisions guide kernel generation and execution scheduling.
The runtime launches the generated kernels and tracks
their execution and completion.

As these representations change, a logical value may acquire
new storage, while different values may share or reuse a region.
The region's contents or ownership can therefore change while
it remains allocated.
Meanwhile, asynchronous device work may still access the region
after the host schedules later operations.
Correct execution depends on preserving which value a region
represents, which operations may access it, and when those
operations complete.
These relations connect compiler representations, generated code,
and runtime state.
XLA's buffer analysis, IREE's compiler--runtime split, and
TorchInductor's generated kernels illustrate these dependencies
\cite{openxlaarchitecture,liu2022tinyiree,ansel2024pytorch2}.
Compiler analyses track aliases and lifetimes to guide storage
decisions, while IR verifiers check local invariants.
Compiler-testing tools compare generated behavior with a reference.
During execution, host and device sanitizers detect supported
memory and synchronization errors
\cite{zhang2025dllibrarytesting,liu2022tzer,yang2024whitefox,
lopes2021alive2,clangasan,nvidiasanitizer}.
Yet a symptom can appear in a different layer from the operation
that lost or misused the required information.
Explaining the failure requires connecting that symptom to
the responsible source decision and its violated condition.

We present \tech{}, an empirical study of tensor binding
failures within and across compiler and runtime layers.
A \emph{tensor execution binding}, or \emph{binding relation},
connects entities that realize a logical tensor value,
including its operations, storage, execution context,
and completion state.
A \emph{binding fact} represents such a relation in concrete
implementation state, such as a storage identity, offset,
alias base, copy policy, placement, or completion state.
A source operation that changes or consumes a binding fact
must satisfy a corresponding \emph{validity condition}.
Section~\ref{sec:binding-model} defines these concepts.

For each repaired cause, \tech{} identifies the
\emph{first invalid decision}: the earliest operation
in the reconstructed causal order whose violated condition
permits the physical effect.
The \emph{observed consequence} describes how that effect
appears in the report.
A \emph{repair} restores validity by changing this decision
or another operation that establishes the required fact
or condition.
Each cause record separates four roles:
\[
  \boxed{\begin{array}{r@{\;:\;}l}
    \text{state} & \text{binding relation and concrete binding fact} \\
    \text{decision} & \text{source operation and its validity condition} \\
    \text{outcome} & \text{physical effect and observed consequence} \\
    \text{change} & \text{landed repair to the fact, operation, or condition}
  \end{array}}
\]
Source tracing connects the binding fact to the reported
physical effect, while comparing pre-fix and repaired revisions
identifies the change that restores validity.
Reports, source revisions, tests, and repairs provide
the supporting evidence.

The study covers IREE, the XLA ecosystem (XLA, JAX, and PJRT),
and the PyTorch ecosystem (PyTorch, TorchInductor, and Triton).
Model-assisted screening and human review select 264 reports
from 7,509 public records.
Of these, 162 have traceable repairs, yielding 149 distinct
repaired causes after deduplication.
Two human reviewers check each cause against its report,
source revisions, and repair.
Three failure families account for 103 of the 149 causes
(69.1\%): \emph{missing eligibility check},
\emph{incorrect mapping or translation}, and
\emph{lost or omitted relation}.
These failures highlight two complementary requirements:
preserving binding facts across representations and enforcing
the validity conditions governing their use.
The required conditions vary with the source operation,
even within the same binding relation.
The first invalid decision occurs in compiler code for
101 causes and runtime code for 48, identifying candidate
checking points in both layers.

At the report level, incorrect result or state is primary
in 168 of 264 reports, motivating output comparisons
and state assertions alongside memory and synchronization checks.
Among four archived IREE diagnostic attempts, two reach the
failing path and detect and localize the physical effect.
One output also identifies the selected invalid-decision site.
Source and repair comparisons establish the violated binding
conditions, distinguishing effect localization from
causal explanation.

This study makes three \textbf{contributions}:
\begin{itemize}[leftmargin=1.25em]
  \item \textbf{Source-linked cause records.}
        We reconstruct repaired causes by connecting binding
        facts, invalid source decisions, validity conditions,
        physical effects, and repairs.
        Each record preserves implementation-specific evidence
        alongside shared taxonomy labels.

  \item \textbf{Cross-stack empirical findings.}
        We characterize 149 distinct repaired causes across
        three compiler and runtime ecosystems.
        The findings identify recurring failure mechanisms,
        transition-specific validity conditions, and candidate
        checking points across compiler and runtime layers.

  \item \textbf{Diagnostic case study.}
        We examine four archived IREE diagnostic attempts
        across five dimensions, from failing-path coverage
        to cause explanation.
        The cases distinguish the diagnostic evidence
        to locate physical effects from the source evidence
        needed to explain their causes.
\end{itemize}

\noindent\textbf{Open science.}
The \artifact{\tech} provides source-linked analysis records,
archived reports and patches, and numerical verification scripts.
The scripts recompute cause-level results from released records
and Phase~1 estimates from reported human-review totals.
Appendix~\ref{app:prompt-execution} describes the available
evidence and replay limitations.

\section{Background and Motivation}\label{sec:background}
AI compiler and runtime stacks progressively map logical tensor values to
physical storage and asynchronous work. This section defines the relations in
that mapping and shows how a state change can invalidate them.

\subsection{From Logical Tensors to Physical Execution}

A tensor in a user program is a logical tensor value. Compilation and
execution can change its physical realization. Compiler transformations may fuse,
split, reorder, or lower tensor operations into executable kernels.
Bufferization and memory planning then assign storage regions, allowing
different logical values to alias or reuse the same region
\cite{li2020dlcompilersurvey,openxlaarchitecture,lattner2021mlir,
ansel2024pytorch2}. 
At runtime, kernels run on streams, queues, or command buffers. Ordering rules,
events, futures, fences, callbacks, or timeline objects represent completion
state
\cite{liu2022tinyiree,nvidiasanitizer}.

A tensor's execution therefore depends on spatial and temporal
relations: which operations and storage realize the value, when those
operations complete, and when the storage can represent a new value.
A binding relation can span compiler and runtime representations.
A compiler can encode operation uses in SSA and storage reuse in buffer
assignments. Lowering can erase those compiler records before execution. The
runtime then represents value--storage, ownership, and completion relations
through allocations, owner records, stream order, and completion objects. A
source trace connects these representations.

\subsection{Tensor Execution Bindings}
\label{sec:binding-model}

Binding relations connect the entities that realize a logical
tensor value: the value, operations, physical storage,
execution context, and completion state.
For example, an operation may depend on earlier work to
complete, while storage has an owner and lifetime.
These relations can be explicit, such as an SSA use--def edge,
or implicit, such as queue ordering.

Binding relations describe different aspects of execution.
Operation relations capture identities and operand roles.
Storage relations capture content, layout, aliasing, ownership,
and lifetime.
Execution-context relations capture device, backend, stream,
and placement, while completion relations capture readiness
and ordering.
A \emph{binding fact} represents a relation in concrete
implementation state, such as a storage identity, offset,
copy policy, or completion state.
The same relation may be represented through IR attributes,
argument indices, generated code, or runtime fields.

Bindings evolve as operations execute.
Allocation, copying, alias creation, and remapping can change
where a value is stored and how operations access it.
Donation, ownership transfer, release, and reuse can change
which operations may access storage and which value it represents.
A \emph{transition} is a source operation that changes or
consumes a binding fact.
Each transition must satisfy a \emph{validity condition},
the source-level requirement governing that operation.
We classify these conditions into normalized families,
called \emph{validity obligations}.
For example, storage reuse requires earlier conflicting
accesses to complete before the region is overwritten.
It must also preserve values still needed through aliases.
Similarly, publishing a new value requires its contents
to be initialized.

\subsection{Binding Failures During Storage Reuse}
\label{sec:motiexample}

Consider two logical tensor values, $v_1$ and $v_2$,
that successively occupy the same physical storage region $m$.
Operation $o_1$ asynchronously reads $v_1$ on execution
context $x_1$, while operation $o_2$ reads $v_2$ after reuse.
The host may schedule this reuse before $o_1$ completes:
\[
  v_1\mapsto m \xrightarrow{\textit{reuse}} v_2\mapsto m.
\]
Valid reuse requires all conflicting accesses by $o_1$
to complete before $m$ is overwritten.
It also requires $v_2$ to be initialized before $o_2$
reads it and any ownership transfer to preserve the
access rights of outstanding operations.
Violating these conditions can cause $o_1$ to read
overwritten data, $o_2$ to read stale data, or an
operation to access storage after its access rights expire.

Four repaired causes illustrate how these conditions fail.
IREE reused a cached allocation for a new parameter
without initializing its contents.
JAX lost a required copy policy during an ownership handoff.
PyTorch performed an indexed update in place without checking
whether another operand shared the destination storage.
Triton moved a shared-memory load past a barrier that
permitted the producer to reuse the region.
Although these failures involve different operations,
each changes or consumes a binding fact without satisfying
its required validity condition.

Understanding these failures requires tracing binding facts,
reconstructing their changes, and identifying the conditions
violated by source operations.
It also requires distinguishing what diagnostics observe
from what source evidence establishes.
These requirements lead to four challenges:

\begin{itemize}[leftmargin=1.25em]
  \item \textbf{C1: Trace binding facts to their consumers.}
        Identify where each binding fact is represented and
        follow its changes to the compiler or runtime decision
        that consumes it.

  \item \textbf{C2: Reconstruct binding evolution.}
        Recover the binding state before and after each transition.
        Determine how translation, resolution, ownership transfer,
        erasure, and reuse change that state.

  \item \textbf{C3: Identify transition-specific validity conditions.}
        Compare pre-fix and repaired paths to identify the
        violated condition and how the repair restores it.
        These conditions include content initialization,
        alias independence, ownership, and completion ordering.

  \item \textbf{C4: Separate diagnostic dimensions.}
        Determine whether a diagnostic invocation reaches
        the failing path, detects and localizes the physical
        effect, and identifies or explains the cause.
        Evaluate these outcomes separately.
\end{itemize}

Section~\ref{sec:study} states the research questions, and
Section~\ref{sec:approach} describes the study method.

\section{Study Scope and Research Questions}
\label{sec:study}

\tech{} studies how tensor bindings fail within and across
compiler and runtime layers, and characterizes their
consequences and repairs.
It also examines what archived diagnostic attempts reveal
about these failures.
The study covers IREE, the XLA ecosystem (XLA, JAX, and PJRT),
and the PyTorch ecosystem (PyTorch, TorchInductor, and Triton).
These stacks use different abstractions for storage planning,
ownership, scheduling, and asynchronous completion
\cite{openxlaarchitecture,jaxdocumentation,liu2022tinyiree,
ansel2024pytorch2,tillet2019triton}.
Their differences allow us to compare how binding relations
are represented and maintained across compiler and runtime layers.

\par\smallskip
\noindent\textbf{Scope boundary.}
The study focuses on failures in the bindings of logical
tensor values already defined by the program.
These bindings govern identity, aliasing, reuse, placement,
representation, ownership, lifetime, dependencies, and handoffs.
A failure is in scope when a compiler or runtime operation
incorrectly establishes, changes, or consumes such a binding.

For example, routing a defined value to the wrong consumer
or operand role violates its binding.
In contrast, a logical computation error is outside the scope
when subsequent storage and execution preserve the resulting
value's bindings.
Unsupported functionality and transformation-local defects
are also excluded when these bindings remain intact.
Three distinctions illustrate this boundary:

\begin{itemize}[leftmargin=1.25em]
  \item \textbf{Logical indices and storage mappings.}
        Losing a view's byte offset maps its elements to an
        incorrect storage range and constitutes a binding failure.
        Computing an incorrect logical index is outside the scope
        when subsequent storage operations use it unchanged.

  \item \textbf{Logical results and external representation.}
        Interpreting a correctly computed unsigned result as signed
        violates its representation binding and is in scope.
        Applying an operation along the wrong axis instead
        produces an incorrect logical result and is outside
        the scope.

  \item \textbf{Compiler consistency and resource ownership.}
        Losing a buffer identity required for release violates
        its ownership binding and is in scope.
        An internal compiler inconsistency is excluded when
        the value's storage and execution bindings remain intact.
        Transformations that lose storage mappings or ownership
        information remain in scope.
\end{itemize}

\par\smallskip
\noindent\textbf{Research questions.}
The six research questions follow the causal explanation
of a binding failure.
RQ1 and RQ2 identify the affected relations and trace their
binding facts to consuming decisions.
RQ3 and RQ4 examine the operations that change or consume
these facts and the conditions governing their validity.
RQ5 characterizes the resulting consequences and repairs,
while RQ6 examines what diagnostic evidence reveals
about the failures.

\begin{itemize}[leftmargin=1.25em]
  \item \textbf{RQ1: Binding relations.}
        Which binding relations are affected by these failures?

  \item \textbf{RQ2: Origins and consumers.}
        Where do the relevant binding facts originate,
        and are they consumed by compiler or runtime decisions?

  \item \textbf{RQ3: Binding transitions.}
        Which operations cause binding failures, and how
        do they change or consume binding facts?

  \item \textbf{RQ4: Validity conditions.}
        Which conditions must these operations satisfy
        to change or consume binding facts correctly?

  \item \textbf{RQ5: Consequences and repair patterns.}
        What consequences are observed?
        For failures with available repairs, which failure
        patterns and repair strategies recur, and how
        do they correspond?

  \item \textbf{RQ6: Diagnostic evidence.}
        What do archived IREE diagnostic attempts reveal
        about failing paths, physical effects, and causes?
\end{itemize}

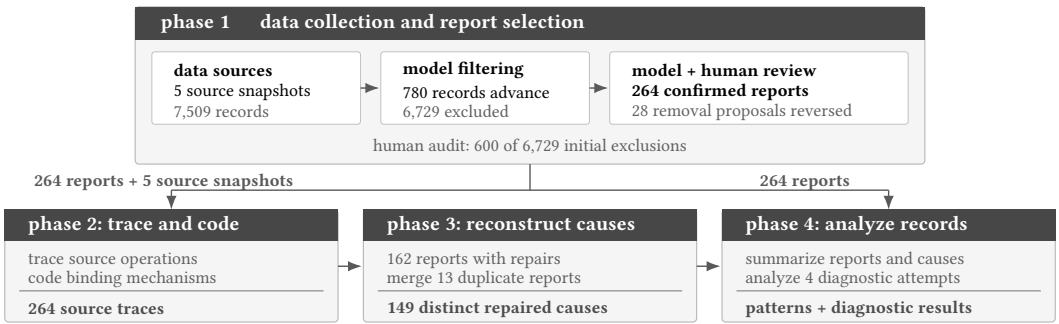
\begin{figure*}[htp]
  \centering
  \begingroup
\setlength{\fboxsep}{0pt}
\resizebox{\textwidth}{!}{%
\begin{tikzpicture}[x=1cm,y=1cm,>=Latex,
  flow/.style={->,draw=black!65,line width=.55pt},
  outer/.style={draw=black!28,fill=black!4,rounded corners=1.5pt,line width=.45pt},
  inner/.style={draw=black!24,fill=white,rounded corners=1.2pt,line width=.4pt}]

  \draw[outer] (1.80,2.18) rectangle (12.70,4.35);
  \fill[black!72] (1.80,3.90) rectangle (12.70,4.35);
  \node[anchor=west,text=white,font=\bfseries\footnotesize]
    at (2.03,4.12) {phase 1};
  \node[anchor=west,text=white,font=\bfseries\footnotesize]
    at (3.40,4.12) {data collection and report selection};

  \draw[inner] (2.03,2.73) rectangle (4.91,3.73);
  \node[anchor=west,font=\bfseries\scriptsize] at (2.21,3.48) {data sources};
  \node[anchor=west,font=\scriptsize] at (2.21,3.18) {5 source snapshots};
  \node[anchor=west,font=\scriptsize,text=black!62] at (2.21,2.90) {7,509 records};

  \draw[inner] (5.19,2.73) rectangle (8.07,3.73);
  \node[anchor=west,font=\bfseries\scriptsize] at (5.37,3.48) {model filtering};
  \node[anchor=west,font=\scriptsize] at (5.37,3.18) {780 records advance};
  \node[anchor=west,font=\scriptsize,text=black!62] at (5.37,2.90) {6,729 excluded};

  \draw[inner] (8.35,2.73) rectangle (12.47,3.73);
  \node[anchor=west,font=\bfseries\scriptsize] at (8.53,3.48) {model + human review};
  \node[anchor=west,font=\bfseries\scriptsize] at (8.53,3.18) {264 confirmed reports};
  \node[anchor=west,font=\scriptsize,text=black!62] at (8.53,2.90) {28 removal proposals reversed};

  \draw[flow] (4.91,3.23) -- (5.17,3.23);
  \draw[flow] (8.07,3.23) -- (8.33,3.23);
  \node[font=\scriptsize,text=black!62] at (7.25,2.43)
    {human audit: 600 of 6,729 initial exclusions};

  \draw[draw=black!65,line width=.55pt] (7.25,2.18) -- (7.25,1.82);
  \draw[draw=black!65,line width=.55pt] (2.30,1.82) -- (12.20,1.82);
  \draw[flow] (2.30,1.82) -- (2.30,1.55);
  \draw[flow] (12.20,1.82) -- (12.20,1.55);
  \node[anchor=west,font=\bfseries\scriptsize,text=black!65]
    at (0.28,1.93) {264 reports + 5 source snapshots};
  \node[anchor=west,font=\bfseries\scriptsize,text=black!65]
    at (10.30,1.93) {264 reports};

  \foreach \x/\w/\phase/\phasetitle/\detail/\detailtwo/\output in {
    0.00/4.60/2/{trace and code}/{trace source operations}/{code binding mechanisms}/{264 source traces},
    4.95/4.60/3/{reconstruct causes}/{162 reports with repairs}/{merge 13 duplicate reports}/{149 distinct repaired causes},
    9.90/4.60/4/{analyze records}/{summarize reports and causes}/{analyze 4 diagnostic attempts}/{patterns + diagnostic results}
  }{
    \draw[outer] (\x,0.00) rectangle ({\x+\w},1.55);
    \fill[black!72] (\x,1.12) rectangle ({\x+\w},1.55);
    \node[anchor=west,text=white,font=\bfseries\footnotesize]
      at ({\x+.20},1.34) {phase \phase: \phasetitle};
    \node[anchor=west,font=\scriptsize,text=black!62]
      at ({\x+.20},.85) {\detail};
    \node[anchor=west,font=\scriptsize,text=black!62]
      at ({\x+.20},.59) {\detailtwo};
    \draw[black!55,line width=.35pt] ({\x+.20},.43) -- ({\x+\w-.20},.43);
    \node[anchor=west,font=\bfseries\scriptsize,text=black!70]
      at ({\x+.20},.19) {\output};
  }
  \draw[flow] (4.60,.77) -- (4.93,.77);
  \draw[flow] (9.55,.77) -- (9.88,.77);
\end{tikzpicture}%
}
\endgroup
  \vspace{-10pt}
  \caption{Study workflow and analysis sets.
  RQ1--RQ4 and cause-level RQ5 use 149 distinct repaired causes.
  Report-level RQ5 uses 264 reports.
  RQ6 examines four archived IREE diagnostic attempts.}
  \ifdefined\Description
    \Description{Phase~1 fixes reference revisions in five
    repositories and screens 7,509 public records.
    Of these, 780 advance to detailed review.
    Model and human review select 264 reports.
    Human review reverses 28 removal proposals; restored reports
    remain in the selected set. Both human reviewers confirm it.
    A separate human audit reviews a random sample of 600
    of the 6,729 initial exclusions.
    Phase~2 traces the selected reports to source operations
    and codes their binding mechanisms.
    Phase~3 compares pre-fix and repaired paths for 162
    repair-linked reports and merges 13 duplicate reports,
    yielding 149 distinct repaired causes.
    Phase~4 analyzes report consequences, failure and repair
    patterns, and four archived IREE diagnostic attempts.}
  \fi
  \label{fig:study-overview}
\end{figure*}

\section{Study Approach}
\label{sec:approach}\label{sec:method}

The study uses three analysis units: \emph{reports},
\emph{distinct repaired causes}, and \emph{diagnostic attempts}.
A report is a public record describing a suspected failure
and serves as the unit of selection and consequence analysis.
A repaired cause is a source-level error supported by a
traceable landed repair.
Reports sharing the same faulty source operation, violated
validity condition, and repair are grouped into one cause
to avoid counting the same error multiple times.
A diagnostic attempt is an archived invocation with recorded
commands and outputs, analyzed against its linked cause.
Figure~\ref{fig:study-overview} presents the four-phase workflow.
Phase~1 collects and reviews public records to select reports
for analysis.
Phase~2 traces binding facts through their representations
to the compiler or runtime decisions that consume them,
addressing C1.
For reports with traceable repairs, Phase~3 compares pre-fix
and repaired paths to reconstruct binding changes and violated
conditions, then merges duplicate reports into distinct causes.
This phase addresses C2 and C3.
Finally, Phase~4 analyzes binding failures, consequences,
and repair patterns across the selected reports and causes.
It also compares four archived IREE diagnostic attempts with
reconstructed causes to address C4.

\subsection{Phase 1: Data Collection and Report Selection}
\label{sec:method:datasource}

Phase~1 establishes versioned source references and selects
public reports for source tracing and consequence analysis.
The process collects candidate records, applies model-assisted
filtering, and validates the selection through human review.

\subsubsection{Data Sources and Collection}

The study covers five repositories across three ecosystems:
IREE, JAX, XLA, PyTorch, and Triton.
A fixed reference revision of each repository provides a
stable starting point for source tracing.
These snapshots locate compiler IR definitions, buffer
assignments, ownership protocols, scheduling, and device
execution.
Individual failures are analyzed at their affected historical
revisions.
For reports with landed repairs, pre-fix and repaired revisions
establish what changed.
Each source reference retains its revision.

\begin{table}[H]
  \centering
  \small
  \caption{GitHub label predicates and unique issue counts.}
  \label{tab:collection-frame}
  \begin{tabular}{@{}
    >{\raggedright\arraybackslash}p{0.24\columnwidth}
    >{\raggedright\arraybackslash}p{0.54\columnwidth}r@{}}
    \toprule
    Repository & Required labels & Issues \\
    \midrule
    \texttt{iree-org/iree} &
      \texttt{bug} (label includes a beetle emoji) & 1,869 \\
    \texttt{pytorch/pytorch} &
      \texttt{module: inductor} $\land$
      \mbox{\texttt{module: correctness (silent)}}
      & 248 \\
    \texttt{triton-lang/triton} &
      \texttt{bug} & 510 \\
    \texttt{jax-ml/jax} &
      \texttt{bug} & 3,908 \\
    \texttt{openxla/xla} &
      \texttt{bug} & 281 \\
    \bottomrule
  \end{tabular}
\end{table}

Candidate records come from GitHub issues and the official
CVE List.
GitHub issues provide engineering discussions and linked
changes, while the CVE List provides security
reports~\cite{cvelistv5}.
Following empirical software engineering sampling
guidance~\cite{baltes2022sampling}, the collection specifies
its search criteria, retrieval procedure, and sampling frame.
The GitHub queries retrieve open and closed issues created
on or before September 24, 2026.
Project-maintained labels define each repository's collection.
Table~\ref{tab:collection-frame} lists the required labels
and unique issue counts.
Each query requires all labels in its row.
For PyTorch, both \texttt{module: inductor} and
\texttt{module: correctness (silent)} are required.
These labels select TorchInductor reports classified as
incorrect results or state without a reported execution error.
Crash, hang, and lifetime reports lacking the
silent-correctness label fall outside this collection.
The other repositories use broader defect labels.
Subsequent screening determines which collected reports
describe binding failures.
Thus, the results characterize these repository-specific
collections rather than the prevalence of defects.
The queries yield 6,816 GitHub issues:
1,869 from IREE, 758 from the PyTorch ecosystem,
and 4,189 from the XLA ecosystem.

CVE collection searches GitHub's code index using six terms:
IREE, PJRT, PyTorch, Triton, XLA, and jaxlib.
The searches identify candidate paths in the official CVE
List, and each returned record is retrieved from the same
fixed revision.
The searches report 713 term-level matches and return
693 paths representing distinct CVE records.
The remaining 20 matches are not retrieved.
The resulting sampling frame contains
\[
  6{,}816\ \text{GitHub issues} + 693\ \text{CVE records}
  = 7{,}509\ \text{candidate records}.
\]

Each record retains its source identifier and selection status.
Appendix~\ref{app:collection-frame} specifies the queries
and collection counts.
The companion artifact provides the query specification,
baseline revisions, record identifiers, and available
capture timestamps.
The issue-creation cutoff determines eligibility, while
capture timestamps record when evidence was retrieved.

\subsubsection{LLM-Assisted Filtering and Human Review}
\label{sec:method:filtering}

Model-assisted filtering reduces the candidate collection
before detailed examination of linked evidence.
Human review of sampled exclusions estimates how often
this step misses binding failures.

Two model reviews separately apply a fixed prompt, codebook,
and output schema to every record.
Each review identifies the logical value, relevant storage
or execution entity, candidate binding relation, reported
violation, and observed consequence.
Supporting quotations connect these decisions to the
record's text.
The reviews cannot access each other's outputs before
completion; a third model review resolves disagreements.
Prior work has also used LLMs for report
screening~\cite{huotala2024llmscreening}.
Each review assigns one of three labels:

\begin{itemize}[leftmargin=1.25em]
  \item \textbf{Include.}
        The record provides evidence of a violated binding
        relation and its observed consequence.

  \item \textbf{Uncertain.}
        The record describes a concrete failure involving
        an in-scope storage or execution relation, but one
        causal link remains unresolved.

  \item \textbf{Exclude.}
        The evidence indicates an out-of-scope cause or
        identifies neither a candidate binding relation
        nor a responsible source operation.
\end{itemize}

Records labeled \emph{include} or \emph{uncertain} advance
to detailed review.
For uncertain records, additional comments, changes, and
tests may resolve the missing causal link.
Exclusion codes distinguish out-of-scope causes from
insufficient evidence.
Appendix~\ref{app:screening-prompt} defines these rules.
After adjudication, 780 GitHub issues advance:
537 labeled \emph{include} and 243 labeled \emph{uncertain}.
The other 6,729 records, including all 693 CVE records,
are excluded.
Before adjudication, the two model reviews agree on
6,387 of 6,816 GitHub issues
(93.71\%; Cohen's $\kappa=0.70$~\cite{cohen1960coefficient})
and 687 of 693 CVE records (99.13\%).
Too few CVE decisions favor advancement for $\kappa$
to be informative.
These values measure agreement between model reviews.

\par\smallskip
\noindent\textbf{Human validation of exclusions.}
The final codebook, model decisions, and review history
are fixed before drawing the audit sample.
The audit draws three non-overlapping random samples of
200 records from the 6,729 exclusions.
Each draw samples uniformly from the remaining records,
so the combined 600-record sample is a simple random
sample without replacement~\cite{cochran1977sampling}.

Two human reviewers independently evaluate all 600 records
using the final codebook.
Their evidence packets contain the archived reports and
available public comments, linked changes, patches, and tests.
The packets hide model decisions and rationales, selection
status, and the other reviewer's labels.
Each reviewer assigns \emph{binding failure} or
\emph{not a binding failure} and cites supporting evidence.
The reviewers then resolve disagreements to establish
the final labels.
The validation tests whether at least 95\% of the filtered
records are correctly excluded.
A one-sided exact 95\% hypergeometric bound accounts for
sampling without replacement from a finite population.
With 600 reviewed records, the criterion is met if at most
21 binding failures are found.
The test has approximately 81\% power when 97\% of the
filtered records are correctly excluded.
All 600 records contribute to the primary
test~\cite{cochran1977sampling,wang2015hypergeometric}.
Table~\ref{tab:filter-validation} reports the individual
samples and combined results.

\begin{table}[htp]
  \centering
  \scriptsize
  \caption{Human validation of the initial exclusions.
  Binding-failure counts use final human labels.
  Reviewer agreement is measured before discussion.}
  \label{tab:filter-validation}
  \setlength{\tabcolsep}{4pt}
  \begin{tabular}{@{}lrrr@{}}
    \toprule
    Group &
    \makecell{Records\\reviewed} &
    \makecell{Binding failures\\found} &
    \makecell{Reviewer agreement\\$n$ (\%)} \\
    \midrule
    Sample 1    & 200 & 0 & 198 (99.0) \\
    Sample 2    & 200 & 3 & 195 (97.5) \\
    Sample 3    & 200 & 1 & 199 (99.5) \\
    All samples & 600 & 4 & 592 (98.7) \\
    \bottomrule
  \end{tabular}
\end{table}

Let $m$ denote the binding failures identified in the sampled
exclusions.
The expansion estimator~\cite{cochran1977sampling} estimates
their total among all 6,729 excluded records:
\[
  \widehat{M}=6{,}729\frac{m}{600}.
\]
The estimated percentage correctly excluded is
$100(1-\widehat{M}/6{,}729)$.
Human review identifies four binding failures, leaving
596 of 600 sampled records (99.33\%) correctly excluded.
With $m=4$, the estimated number of missed failures is 45.
The equal-tailed exact 95\% hypergeometric interval is
14--111, corresponding to 98.35\%--99.79\% correctly
excluded.
The one-sided 95\% lower bound is 98.53\%, exceeding the
95\% criterion~\cite{wang2015hypergeometric}.
These estimates use the reported human-review totals.
The public release does not include the sample membership
lists or individual human decisions.
Its scripts therefore verify the calculations but cannot
independently reproduce the sample draws or review labels.

\subsubsection{Detailed Review and Final Selection}

Detailed review examines linked evidence to resolve causal
gaps in the 780 records that pass filtering.
Each evidence packet includes the report's public comments,
timeline, cross-referenced issues, pull requests, patches,
and tests.
These materials connect the observed symptom to the
implementation and any proposed or incorporated repair.

The model reviews follow the initial screening procedure
and classify each record as \emph{confirmed},
\emph{unresolved}, or \emph{excluded}.
At each step, two human reviewers independently examine
all proposed removals, including unresolved records.
They record their judgments before resolving disagreements.
Reports confirmed as binding failures return to the
candidate set before the next step.
Detailed review is followed by four additional rounds
and a final review.
Codebook revisions are reapplied to all 780 advanced records.
Table~\ref{tab:selection-flow} records each step's input,
proposed removals, human restorations, final removals,
and output.
Each output becomes the next step's input.

\begin{table}[htp]
  \centering
  \scriptsize
  \caption{Report selection after human review at each step.
  Restored reports remain in the output set.}
  \label{tab:selection-flow}
  \setlength{\tabcolsep}{4pt}
  \begin{tabular}{@{}lrrrrr@{}}
    \toprule
    Review step & Input &
    \makecell{Proposed\\removals} & Restored &
    \makecell{Final\\removals} & Output \\
    \midrule
    Detailed review
      & 780 & 366 & 19 & 347 & 433 \\
    Review round 1
      & 433 & 41 & 3 & 38 & 395 \\
    Review round 2
      & 395 & 67 & 5 & 62 & 333 \\
    Review round 3
      & 333 & 55 & 1 & 54 & 279 \\
    Review round 4
      & 279 & 10 & 0 & 10 & 269 \\
    Final review
      & 269 & 5 & 0 & 5 & 264 \\
    \midrule
    All steps
      & -- & 544 & 28 & 516 & 264 \\
    \bottomrule
  \end{tabular}
\end{table}

Human review reverses 28 of 544 proposed removals.
The remaining 516 removals leave 264 reports, including
the restored reports.
The 28 reversals count review decisions across steps;
the released aggregate log cannot establish whether a
record was restored more than once.
Both human reviewers independently apply the final codebook
to all 264 retained reports.
Before discussion, both classify every report as a binding
failure.
Final-set precision relative to these human labels is
therefore $264/264=100\%$~\cite{manning2008ir}.
The companion artifact records final report membership
and the reported review totals.
Per-step human decisions and restored-record identifiers
are unavailable in the public release.
All 264 reports proceed to source tracing.
Reports with traceable landed repairs also enter cause
reconstruction and deduplication.

\subsubsection{Estimating Initial Filtering Recall}

Recall measures the fraction of evidence-supported binding
failures retained during initial filtering.
A positive reference label requires public evidence of a
binding relation, its violation, and an observed consequence.
Records lacking evidence for any required element receive
a negative label.
Human review confirms 264 binding-failure reports among
the 780 records that pass initial filtering.
The audit estimates the number missed among the initial
exclusions.
Using $\widehat{M}=44.86$ from
Section~\ref{sec:method:filtering}, estimated recall
is~\cite{manning2008ir}
\[
  \widehat{R}_{\mathrm{initial}}
    =\frac{264}{264+\widehat{M}}
    =\frac{264}{264+44.86}
    =85.48\%.
\]
Applying the same calculation to the 95\% interval of
14--111 missed failures gives a recall interval of
70.40\%--94.96\%.
This estimate applies to evidence-supported failures
within the collected sampling frame.
The four positive audit records contribute to the
estimated omissions among the 6,729 initial exclusions;
they remain separate from the 264 analysis set.
Appendix~\ref{app:selection-audit} reports the review counts.

\subsection{Phase 2: Source Tracing and Binding Coding}
\label{sec:method:binding}

Phase~2 traces the 264 human-confirmed reports to candidate
source operations.
Each trace connects the reported physical effect to the binding
facts, transitions, and validity conditions that explain it.

\subsubsection{Trace Scope}

Tracing begins at the reported effect or a linked source location.
Production and consumption traces follow operands, results,
use--def edges, and scheduler users.
Reuse and ownership traces follow aliases, assignments, retains,
releases, moves, and donations through the last affected use.
For asynchronous execution and transfers, traces follow launches,
completion objects, dependencies, source and destination storage,
and execution contexts.
A trace may span IR attributes, generated code, runtime objects,
and scheduler state.

\subsubsection{Source Tracing}
\label{sec:method:source-tracing}

Source tracing identifies the earliest operation whose violated
validity condition permits the reported physical effect.
Algorithm~\ref{alg:source-tracing} summarizes this analyst-guided
procedure.
It produces a cited trace, the earliest supported candidates,
and either a selected decision or an unresolved status.

\begingroup
\SetAlCapFnt{\footnotesize}
\SetAlCapNameFnt{\footnotesize\bfseries}
\begin{figure}[htp]
\centering
\begin{minipage}{0.75\linewidth}

\begin{algorithm}[H]
\footnotesize
\setlength{\algomargin}{1.2em}
\DontPrintSemicolon
\LinesNumbered
\SetAlgoNlRelativeSize{-1}
\caption{Trace a report to the first invalid decision}
\label{alg:source-tracing}

\SetKwFunction{Anchors}{linkedAnchors}
\SetKwFunction{Locate}{locateAnchors}
\SetKwFunction{Resolve}{resolveHistoricalLocations}
\SetKwFunction{Backward}{traceBackward}
\SetKwFunction{Forward}{traceForward}
\SetKwFunction{Cite}{retainCitedSteps}
\SetKwFunction{Candidates}{candidateDecisions}
\SetKwFunction{Check}{retainSupported}
\SetKwFunction{Order}{documentedOrder}
\SetKw{Return}{return}

\KwIn{$r$: report; $E$: linked evidence;
      $S$: versioned source snapshots}
\KwOut{$T$: cited trace; $C$: earliest supported candidates;
       $d$: first invalid decision or $\bot$}

$A\gets\Anchors(r,E)$
\nllabel{alg:trace:anchor}
\tcp*[r]{\scriptsize\textcolor{green!45!black}{linked source locations}}

\If{$A=\emptyset$}{
    $A\gets\Locate(r,S)$
    \tcp*[r]{\scriptsize\textcolor{green!45!black}{search snapshots}}
}
\If{$A=\emptyset$}{
    \Return{$(\emptyset,\emptyset,\bot)$}\;
}

$H\gets\Resolve(r,A,E,S)$
\nllabel{alg:trace:revision}
\tcp*[r]{\scriptsize\textcolor{green!45!black}{affected historical code}}

\If{$H=\emptyset$}{
    \Return{$(\emptyset,\emptyset,\bot)$}\;
}

\BlankLine
$T\gets\emptyset$\;
\ForEach{$h\in H$}{
    $T\gets T\cup\Backward(h,E,S)$
    \nllabel{alg:trace:backward}
    \tcp*[r]{\scriptsize\textcolor{green!45!black}{trace to the value}}

    $T\gets T\cup\Forward(h,E,S)$
    \nllabel{alg:trace:forward}
    \tcp*[r]{\scriptsize\textcolor{green!45!black}{trace to the effect}}
}

$T\gets\Cite(T,E,S)$
\nllabel{alg:trace:evidence}
\tcp*[r]{\scriptsize\textcolor{green!45!black}{retain cited steps}}

\BlankLine
$D\gets\Candidates(T,E)$
\nllabel{alg:trace:candidates}
\tcp*[r]{\scriptsize\textcolor{green!45!black}{candidate violations}}

$D\gets\Check(D,T,E)$
\nllabel{alg:trace:check}
\tcp*[r]{\scriptsize\textcolor{green!45!black}{require causal evidence}}

$\prec\;\gets\Order(D,T,E)$
\nllabel{alg:trace:order}
\tcp*[r]{\scriptsize\textcolor{green!45!black}{documented causal order}}

$C\gets\{c\in D\mid\not\exists c'\in D: c'\prec c\}$
\nllabel{alg:trace:earliest}
\tcp*[r]{\scriptsize\textcolor{green!45!black}{no earlier candidate}}

$d\gets\bot$\;
\If{$C=\{c\}$}{
    $d\gets c$
    \nllabel{alg:trace:select}
    \tcp*[r]{\scriptsize\textcolor{green!45!black}{unique earliest decision}}
}

\Return{$(T,C,d)$}\;

\end{algorithm}
\end{minipage}
\ifdefined\Description
  \Description{Analysts locate the affected historical code,
  trace binding information, and retain cited steps.
  They select the first invalid decision when the documented
  causal order leaves one earliest supported candidate.
  Otherwise, the report remains unresolved.}
\fi
\end{figure}
\endgroup

Analysts first locate source anchors and resolve them to the
affected historical revisions.
When no source anchor is linked, reference snapshots provide
a starting point
(lines~\ref{alg:trace:anchor}--\ref{alg:trace:revision}).
Backward tracing follows each binding fact to its documented
origin: an API value, IR fact, runtime field, or backend contract.
Forward tracing follows the fact to the reported physical effect.
The trace retains each supported operation that creates, changes,
drops, or consumes the fact, together with its cited evidence
(lines~\ref{alg:trace:backward}--\ref{alg:trace:evidence}).
A candidate changes or consumes a binding fact while its
validity condition is false.
Source evidence must establish this violation and connect
the operation to the physical effect
(lines~\ref{alg:trace:candidates}--\ref{alg:trace:check}).
A competing access remains in the trace but is excluded
as a candidate when evidence establishes that it follows
its documented contract.

Each candidate is the smallest source operation supported
by the evidence that makes the binding decision on the
failing path.
Examples include an assignment that publishes state,
a branch that permits a transformation, a helper call
that performs a transfer, or an IR rewrite that emits
an operation.
When the evidence identifies only a helper, the trace
records its call.
For a lost binding fact, the candidate is the operation
that drops information its contract requires it to preserve.
An omitted action has no executed source location.
In this case, the candidate is the first executed operation
that requires the missing result, while the failure family
remains \emph{omitted operation}.
An earlier operation replaces a later candidate only when
evidence connects its violation to the same physical effect.

After identifying supported violations, the analysis orders
candidates using program order and data or control
dependencies within an execution path.
Documented synchronization or happens-before edges establish
the order between concurrent operations.
Across compilation and execution, the causal order follows
the changed IR, generated code, or launch configuration
from a compiler decision to the runtime operation that uses it.
Candidates without an earlier supported candidate form $C$
(lines~\ref{alg:trace:order}--\ref{alg:trace:earliest}).
A unique candidate becomes the selected decision $d$.
Otherwise, $d=\bot$, and the report remains unresolved
for Phase~3.

\subsubsection{Coding Relations, Transitions, and Obligations}

Each trace records four fields that distinguish the affected
relation, its provenance, the source operation, and its
validity condition:

\begin{itemize}[leftmargin=1.25em]
  \item \textbf{Relation.}
        This field identifies the entities connected at the
        first invalid decision.
        The primary relation is the one whose binding fact
        the decision changes or consumes and the repair
        restores or checks.
        Other supported relations remain in the trace.

  \item \textbf{Provenance path.}
        This field traces a binding fact from its origin
        to the first invalid decision.
        The path may include translation, validation, resolution,
        ownership transfer, or code emission.

  \item \textbf{Transition.}
        This field records the source operation, binding states
        before and after it, changed fields, governing conditions,
        trigger, and physical effect.

    \item \textbf{Validity obligation.}
      The codebook maps source-level validity conditions
      to normalized obligation families.
      Operation selection precedes condition coding.
      The primary condition is the requirement
      violated at the first invalid decision.
      The repair restores or enforces this requirement,
      either at that decision or at another operation
      that establishes the necessary binding state.
      The corresponding codebook family is assigned
      as the primary obligation.
      A regression test or an attributed developer
      diagnosis may help identify the requirement
      restored by the repair.

\end{itemize}

The pre-fix/repaired comparison provides evidence for selecting
the source operation and its primary condition.
If either remains unresolved, no cause record is produced.
Each coded label retains its implementation-specific
term and source location.
A binding fact originates where it first acquires a concrete
representation.
Origins fall into four categories: API or frontend, compiler
IR or analysis, runtime state, and backend or device contract.
Its consumer is the compiler or runtime decision that drops,
changes, or consumes the fact.
Each repaired cause contributes one primary provenance record
to RQ2.
This record follows the fact required to evaluate the primary
validity condition.
Related values, such as a base address and offset, share
the same record.
RQ2 counts the fact's origin and the first invalid
decision as provenance endpoints.
The complete source trace also retains intermediate steps
and the downstream physical effect.

The research team checks each candidate's causal path,
validity condition, and labels against the cited report
and source.
When analysts disagree, they revisit the supporting
definition--use, call, and synchronization links.
Supported alternatives remain available for the Phase~3
repair comparison.
Each trace links its coded fields to report identifiers,
source locations, revisions, and supporting evidence.

\subsection{Phase 3: Cause Reconstruction and Deduplication}
\label{sec:method:cases}

Phase~3 reconstructs binding failures by comparing pre-fix
and repaired source paths.
The comparison identifies the violated conditions and
explains how the repairs restore validity.
Coding, deduplication, and human review produce the final
cause records.
Of the 264 source-tracing reports, 162 have traceable repairs
and enter reconstruction.
The remaining 102 retain their reviewed report-level records.

\begingroup
\SetAlCapFnt{\footnotesize}
\SetAlCapNameFnt{\footnotesize\bfseries}
\begin{figure}[htp]
\centering
\begin{minipage}{0.75\linewidth}

\begin{algorithm}[H]
\footnotesize
\setlength{\algomargin}{1.2em}
\DontPrintSemicolon
\LinesNumbered
\SetAlgoNlRelativeSize{-1}
\caption{Reconstruct a cause from its repair}
\label{alg:cause-reconstruction}

\SetKwFunction{First}{first}
\SetKwFunction{Last}{last}
\SetKwFunction{Parent}{parent}
\SetKwFunction{Sites}{repairSites}
\SetKwFunction{Trace}{reconstructPath}
\SetKwFunction{Compare}{comparePaths}
\SetKwFunction{Decision}{resolveDecision}
\SetKwFunction{Condition}{primaryCondition}
\SetKwFunction{Update}{updateTrace}
\SetKwFunction{Build}{buildCauseRecord}
\SetKwFunction{Cite}{linkEvidence}
\SetKw{Return}{return}

\KwIn{$T$: Phase~2 trace; $E$: archived evidence;
      $P$: ordered repair commits}
\KwOut{$K$: cause record, or $\bot$ if unresolved}

$\nu^{-}\gets\Parent(\First(P))$
\nllabel{alg:cause:pre}
\tcp*[r]{\scriptsize\textcolor{green!45!black}{pre-fix revision}}

$\nu^{+}\gets\Last(P)$
\nllabel{alg:cause:post}
\tcp*[r]{\scriptsize\textcolor{green!45!black}{repaired revision}}

$L\gets\Sites(P,E)$
\tcp*[r]{\scriptsize\textcolor{green!45!black}{repair and patch sites}}

\BlankLine
$T^{-}\gets\Trace(T,L,\nu^{-},E)$
\nllabel{alg:cause:paths}\;
$T^{+}\gets\Trace(T,L,\nu^{+},E)$\;

$\Delta\gets\Compare(T^{-},T^{+})$
\tcp*[r]{\scriptsize\textcolor{green!45!black}{binding changes}}

\BlankLine
$d\gets\Decision(T^{-},\Delta,E)$
\nllabel{alg:cause:decision}
\tcp*[r]{\scriptsize\textcolor{green!45!black}{earliest violation}}

\If{$d=\bot$}{
    \Return{$\bot$}\;
}

$o\gets\Condition(d,\Delta,E)$
\nllabel{alg:cause:condition}
\tcp*[r]{\scriptsize{condition at $d$}}

\If{$o=\bot$}{
    \Return{$\bot$}\;
}

\BlankLine
$T\gets\Update(T,d,E)$
\nllabel{alg:cause:update}
\tcp*[r]{\scriptsize\textcolor{green!45!black}{update upstream relations}}

$K\gets\Build(T,d,o,L,\Delta,P,E)$
\nllabel{alg:cause:record}\;

$K\gets\Cite(K,E)$
\tcp*[r]{\scriptsize\textcolor{green!45!black}{support each claim}}

\Return{$K$}\;

\end{algorithm}
\end{minipage}
\ifdefined\Description
  \Description{Analysts compare pre-fix and repaired paths
  to identify changed binding facts, operations, and conditions.
  They resolve the first invalid decision and primary condition,
  update the source trace, and build a cited cause record.
  Unresolved cases return no cause record.}
\fi
\end{figure}
\endgroup

\subsubsection{Cause Reconstruction}

Each reconstruction combines the Phase~2 trace, archived
report evidence, and a repair incorporated into the project's
repository.
Comparing pre-fix and repaired paths identifies changes
to binding facts, operations, and validity conditions.
The record distinguishes three source locations.
The \emph{first invalid decision} introduces the violation.
The \emph{repair site} is the operation whose behavior changes,
while the \emph{patch site} is the edited source location.
These locations may differ.
Linking the edited code to the changed operation explains
how the repair addresses the original violation.

Algorithm~\ref{alg:cause-reconstruction} summarizes the
analyst-guided procedure.
Analysts first identify the pre-fix and repaired revisions
(lines~\ref{alg:cause:pre}--\ref{alg:cause:post}).
Both paths record the same fields: logical value, operation,
storage, execution context, completion state, transition,
and physical effect.
Comparing these fields resolves the candidate decision
from Phase~2 and identifies the validity condition
violated at that decision
(lines~\ref{alg:cause:paths}--\ref{alg:cause:condition}).
If either remains unresolved, no cause record is produced.
If the selected decision changes, the trace and its upstream
relations are updated (line~\ref{alg:cause:update}).

The repair identifies what changed, while tracing beyond
the edited lines connects that change to the original violation.
For a multi-hunk patch, the repair strategy follows the
operation or condition changed to restore validity,
rather than the number of edited locations.
Each record distinguishes the reported release or branch
from the revisions used for comparison.
It also records the availability of patches,
source comparisons, regression tests, attributed developer
diagnoses, and public reproducers.

\subsubsection{Coding, Deduplication, and Human Review}

The cause record extends the Phase~2 record with four fields.
The \emph{failure family} describes how the first invalid
decision violates its validity condition.
The \emph{physical effect} captures the resulting execution
behavior, while the \emph{observed consequence} describes
how that effect appears in the report.
The \emph{repair strategy} describes how the change at the
repair site restores or enforces the violated condition.
For example, an incorrect offset may cause an out-of-range
access that appears as a crash.
Later invalid operations remain in the causal path.

Coding and review involve four tasks:

\begin{enumerate}[leftmargin=1.5em,itemsep=0.5em]
  \item \textbf{Code consequences.}
        Each report receives one primary consequence:
        incorrect result or state, runtime rejection or crash,
        memory-access violation, data race, hang or deadlock,
        resource leak or exhaustion, or another observable failure.
        Three nonexclusive flags capture memory-access signals,
        concurrent-access violations, and availability loss.
        A security consequence requires cited evidence linking
        the effect to a protected asset and establishing the
        attacker capability and system conditions required
        for that impact.

  \item \textbf{Build shared categories.}
        The binding model provides the initial relation categories.
        Source descriptions and repairs supply candidate
        transition, obligation, failure, and repair labels.
        Analysts merge labels that connect the same entity types
        and satisfy the same definition.
        Categories are split when their state changes,
        validity conditions, failure mechanisms, or repairs differ.
        Each codebook revision is reapplied to all causes.
        Records retain their implementation-specific terms
        alongside the shared categories.

  \item \textbf{Merge duplicate reports.}
        Reports share a cause identifier only when their
        source operation, primary condition, and repair
        behavior match.
        This rule also merges backports whose edited lines differ.
        Distinct invalid decisions remain separate causes,
        even when one patch repairs them.
        Reports with different primary conditions also remain
        separate, even when they involve the same source operation.

  \item \textbf{Validate the records.}
        Following prior compiler-bug
        studies~\cite{du2021compilerbugs,shen2021compilerbugs},
        two human reviewers check every cause against its
        report, source revisions, tests, and repair.
        They examine the causal explanation, coded fields,
        provenance endpoints, and duplicate links.
        Fields inconsistent with the evidence are corrected
        and rechecked.
        Conflicting developer explanations remain archived
        and are resolved through source comparison.
        The final report reviews provide the
        remaining consequence labels.
\end{enumerate}

The final codebook contains eight relation, seven transition,
15 obligation, and eight failure families, plus seven repair
strategies.
The results use the frozen 2026-10-01 codebook.
Appendix~\ref{app:taxonomy-codebook} defines these families
and their boundaries.
Source comparison resolves all 162 repair-linked reports.
Merging 13 duplicate reports yields 149 distinct repaired
causes, each with one first invalid decision and one
primary obligation.
Both human reviewers agree on the final cause records
and report-to-cause mapping.
Independent source-operation selections were not recorded
before reconciliation.
These records support RQ1--RQ4 and cause-level RQ5.
The companion artifact retains the supporting evidence
and report-to-cause links.

\subsubsection{Worked Examples}
\label{sec:method:worked-examples}

\begin{figure}[H]
  \centering
  \includegraphics[width=\linewidth]{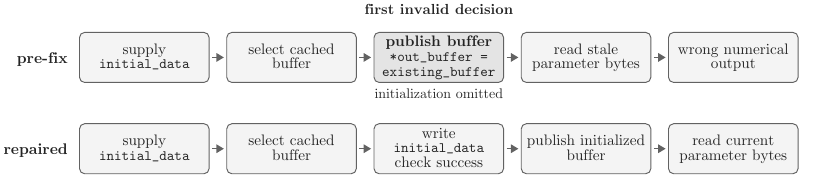}
  \vspace{-15pt}
  \caption{Pre-fix and repaired paths for buffer reuse.
  The repair writes \texttt{initial\_data} and checks success
  before publishing the buffer.}
  \Description{Both paths receive new parameter bytes and select
  a cached buffer. The pre-fix path publishes the buffer without
  writing those bytes, so later reads use stale contents and
  produce wrong numerical output. The repaired path writes
  the bytes, checks success, and publishes the initialized buffer.}
  \label{fig:worked-initialization}
\end{figure}

IREE's cached-buffer report~\cite{tb-iree-12161}
describes incorrect numerical output after repeated parameter updates.
Each update supplies new bytes through \texttt{initial\_data}.
In the pre-fix reuse branch of
\texttt{iree\_hal\_caching\_allocator\_pool\_acquire},
\texttt{*out\_buffer=existing\_buffer} publishes a cached
buffer without writing the new bytes.
Figure~\ref{fig:worked-initialization} compares the two paths.

\noindent\textbf{Decision selection.}
The omitted write has no executed source location.
The publication assignment is the first invalid decision
because it precedes the return and downstream read without
initializing the contents.
The binding relation connects the logical parameter to
storage content, the transition is \emph{initialization or
content update}, and the primary obligation is
\emph{initialization before publication}.

\noindent\textbf{Repair and evidence.}
The repair inserts \texttt{iree\_hal\_buffer\_map\_write}
before publication, checks its success, and releases the
buffer on failure.
The source comparison, maintainer diagnosis, reproducer,
and reporter retest support this reconstruction.
The failure family is \emph{omitted operation}, repaired by
performing the missing write.
Stale parameter content causes the incorrect output.

\vspace{3pt}
\noindent\textbf{Concurrent access to semaphore identity.}
In IREE's semaphore-cleanup case~\cite{tb-iree-9392},
callback processing clears the base timepoint under
\texttt{timepoint\_mutex}, as required before handler invocation.
Waiter cleanup reads \texttt{timepoints[i].base.semaphore}
outside the lock to obtain an argument for a cancellation
function that later acquires the same mutex.
This read can race with the callback reset.\footnote{
Pre-fix revision \texttt{d42e002e}:
\texttt{semaphore\_base.c}, lines~75--88, 121--162, 208--233,
and 238--268; \texttt{task\_semaphore.c},
lines~30--35 and 391--397.}
The reset follows the documented lifecycle, while the
cleanup read depends on an identity that the callback can clear.
The maintainer diagnosis identifies this lookup as the error.
Neither access precedes the other in the happens-before order,
making the cleanup read the sole supported invalid decision.
Its primary obligation is \emph{happens-before ordering}.
The caller retains the semaphore, and timepoint storage
remains allocated through cleanup.
The repair saves the semaphore identity in a separate field
at acquisition and uses it during cleanup.
The callback still resets the base, but no longer conflicts
with the cleanup read.
The repair \emph{preserves or propagates the relation}
without changing semaphore reference counting.

\noindent\textbf{Alias independence in scatter reinplacement.}
\par\nobreak\noindent
In PyTorch's scatter-reinplacement case~\cite{tb-pytorch-197829},
\texttt{should\_reinplace\_scatter} replaces a functional
scatter with \texttt{view(inp).copy\_(src)}.
This transformation can change the result when
\texttt{inp} and \texttt{src} share storage.
The repair checks their storage identities and retains the
functional scatter when they alias, enforcing the primary
obligation of \emph{alias independence}.
\par

\subsection{Phase 4: Cross-Case Analysis and Exploratory Diagnostic Case Study}
\label{sec:method:detection}

Phase~4 aggregates the coded records for RQ1--RQ5 and compares
archived diagnostic attempts with reconstructed causes for RQ6.
For RQ1--RQ4, each repaired cause contributes one primary
relation, provenance record, transition, and validity obligation.
To examine how validity conditions vary within a binding relation,
we count primary relation--transition--obligation combinations
across the 149 causes.
RQ5 pairs failure families with repair strategies at the cause
level and counts consequences across the reports selected in Phase~1.
RQ6 examines four archived attempts from three IREE causes.
Each attempt has preserved commands and outputs.
Its record also retains the tool or check, configuration,
input, executed path, reported location, and linked cause.
Analysts align the invocation's backend and executed path
with the source operation and first invalid decision
in the cause record.
Diagnostic output remains separate from later developer diagnoses
so that the assessment reflects what the diagnostic itself reports.

Five dimensions distinguish reaching the failing path,
detecting its effect, and locating and explaining its cause:

\begin{itemize}[leftmargin=1.25em,itemsep=0.2em]
  \item \textbf{Failing-path coverage.}
        Does it reach the first invalid decision
        on the failing path?

  \item \textbf{Effect detection.}
        Does it signal the physical effect?

  \item \textbf{Effect localization.}
        Does it identify the concrete runtime object,
        allocation, access, or source site where the
        physical effect appears?

  \item \textbf{Cause localization.}
        Does it identify the exact source operation selected
        as the first invalid decision?

  \item \textbf{Cause explanation.}
        Does it explain the binding relation or validity
        condition that makes the decision invalid?
\end{itemize}

Effect localization can identify an affected allocation or
conflicting access; cause localization requires the selected
source operation itself.
Each dimension receives one of five codes: \emph{yes},
\emph{no}, \emph{unreported}, \emph{not evaluated},
or \emph{not applicable}.
When an attempt does not reach the first invalid decision,
the other four dimensions are \emph{not evaluated}.
This separates a path mismatch from a detection failure.
For \emph{no} or \emph{unreported}, the record states the reason:
a path mismatch, a missing binding fact, an available fact
left unchecked, or the operation where the diagnostic stops.
Each code cites the archived output and source evidence.
The linked repair record identifies the required binding fact.
The released RQ tables link counts to report, cause,
or attempt identifiers. Scripts check their membership,
distributions, and evidence-file hashes.

\section{Empirical Results}\label{sec:eval}

\noindent\textbf{Study setup.}
The released analysis scripts require Python 3.10 or later.
We verified the archived analysis on Ubuntu 24.04.3 LTS
with Python 3.12.3.
The model reviews used Codex CLI 0.155.0,
the model alias \texttt{gpt-daybreak-blue-latest},
and high reasoning effort.
Appendices~\ref{app:screening-prompt}
and~\ref{app:validation-prompt} summarize the review rules;
Appendix~\ref{app:prompt-execution} documents execution
settings and output validation.
Source tracing uses the fixed snapshots and historical
revisions described in Phase~1.
RQ6 analyzes archived diagnostic commands and outputs
under their recorded configurations.

\vspace{3pt}
\noindent\textbf{Analysis sets.}
The evaluation uses three analysis units:
reports for consequence analysis, distinct repaired causes
for RQ1--RQ4 and cause-level RQ5, and diagnostic attempts
for RQ6.
Table~\ref{tab:analysis-sets} summarizes these sets.
The study includes 264 human-confirmed reports:
74 from IREE, 99 from the PyTorch ecosystem, and 91
from the XLA ecosystem.
Of these, 162 have traceable repairs.
Merging 13 reports that share a first invalid decision
and repair with another report yields 149 distinct
repaired causes: 46 from IREE, 55 from the PyTorch
ecosystem, and 48 from the XLA ecosystem.
The remaining 102 reports contribute only to
report-level analysis.

\begin{table}[H]
  \centering
  \small
  \caption{Analysis sets, repair availability, and analytical uses.}
  \vspace{-5pt}
  \label{tab:analysis-sets}
  \setlength{\tabcolsep}{4pt}
  \begin{tabular}{@{}
    >{\raggedright\arraybackslash}p{.35\linewidth}
    r
    >{\raggedright\arraybackslash}p{.43\linewidth}@{}}
    \toprule
    Set & Count & Analytical use \\
    \midrule
    Source-tracing reports
      & 264
      & Source tracing and report-level consequence analysis \\

    Reports with traceable repairs
      & 162
      & Pre-fix/repaired comparison and cause reconstruction \\

    Reports without traceable repairs
      & 102
      & Report-level consequence analysis only \\

    Distinct repaired causes
      & 149
      & RQ1--RQ4 and cause-level RQ5 \\

    Reports with consequence codes
      & 264
      & Report-level RQ5 \\

    Diagnostic attempts
      & 4
      & RQ6; archived attempts from three IREE causes \\
    \bottomrule
  \end{tabular}
\end{table}

\subsection{RQ1: Binding Relations}\label{sec:rq1}

Value--storage/layout is the largest primary relation among
the 149 repaired causes, accounting for 37 (24.8\%).
The other 112 causes (75.2\%) span seven families
(Table~\ref{tab:rq1-bindings}).
These cover operation identity, execution context, completion,
ownership, shared storage, representation, and content.

\begin{table}[htp]
  \centering
  \caption{Primary binding relations in 149 repaired causes.}
  \vspace{-5pt}
  \label{tab:rq1-bindings}
  \small
  \renewcommand{\arraystretch}{1.08}
  \begin{tabular}{@{}lrrrr@{}}
    \toprule
    Relation family & IREE & \makecell{PyTorch\\ecosystem} &
    \makecell{XLA\\ecosystem} & Total \\
    \midrule
    Value--storage/layout           & 10 & 21 &  6 & 37 \\
    Value--operation/identity       & 10 &  8 &  5 & 23 \\
    Operation--execution context    &  6 &  0 & 16 & 22 \\
    Operation--completion           &  4 &  9 &  8 & 21 \\
    Storage--ownership/lifetime     &  9 &  0 &  7 & 16 \\
    Value--shared storage           &  3 &  9 &  2 & 14 \\
    Value--representation           &  2 &  8 &  3 & 13 \\
    Logical value--storage content  &  2 &  0 &  1 &  3 \\
    \midrule
    Total                           & 46 & 55 & 48 & 149 \\
    \bottomrule
  \end{tabular}
\end{table}

The leading relation differs across the sampled ecosystems.
Storage/layout leads in the PyTorch ecosystem (21 of 55 causes),
followed by completion and shared storage (nine each).
Execution context leads in the XLA ecosystem (16 of 48),
followed by completion (8) and ownership/lifetime (7).
In IREE, storage/layout and operation identity each account
for ten of 46 causes, followed by ownership/lifetime (9).
Beyond storage layout, 58 causes concern operation identity,
representation, or execution context.
The 23 operation-identity causes involve argument ordinals,
result positions, binding identifiers, and coordinate maps.
For example, an IREE implementation used sparse buffer
identifiers as indices into a dense kernel-argument array.
This could select an absent or incorrect slot.
The repair established a consistent dense mapping across
compiler and runtime representations.
The other two families concern how values are interpreted
and where their operations execute.
The 13 representation causes involve signedness, sub-byte
storage, and precision configuration.
The 22 execution-context causes involve device commitment,
sharding, streams, and workgroup placement.
These cases show that the affected relation can govern
the use of a value rather than its storage layout.

{Completion relations take different forms across stacks.}
In the analyzed cases, IREE represents completion through
timepoints and task notifications.
The PyTorch ecosystem uses events and barriers, while the
XLA ecosystem uses tokens, callbacks, or collective state.

\find{Storage/layout is primary in 37 of 149 repaired causes
(24.8\%); the other 112 span seven relation families.
This diversity motivates relation-specific checks grounded
in each stack's concrete representations.}

\subsection{RQ2: Provenance Endpoints}\label{sec:rq2}

Compiler decisions consume the primary binding fact in 101 of
the 149 repaired causes, while runtime decisions consume it
in 48 (Table~\ref{tab:rq2-layers}).
API/frontend state and compiler IR/analysis are the two largest
origin categories, contributing 64 facts each.
The remaining facts originate in runtime state or backend
and device contracts.

\begin{table}[ht]
  \centering
  \caption{Origins and consuming layers of 149 primary
  binding facts.}
  \vspace{-5pt}
  \label{tab:rq2-layers}
  \small
  \renewcommand{\arraystretch}{1.05}
  \begin{tabular}{@{}lrrr@{}}
    \toprule
    Fact origin & Compiler decision & Runtime decision & Total \\
    \midrule
    API or frontend             & 41 & 23 & 64 \\
    Compiler IR or analysis     & 59 &  5 & 64 \\
    Runtime state               &  0 & 19 & 19 \\
    Backend or device contract  &  1 &  1 &  2 \\
    \midrule
    Total                       & 101 & 48 & 149 \\
    \bottomrule
  \end{tabular}
\end{table}

Origin and consumer categories differ in 71 records.
These comprise all 64 API/frontend origins, five compiler-origin
facts consumed at runtime, and two backend or device-contract
origins.
Of the API/frontend facts, 41 reach compiler decisions and
23 reach runtime decisions.
This count compares endpoints, not intermediate layer crossings.
The source traces show how binding facts change representation
before reaching these decisions.
Tensor views become storage offsets, API requests become
transfer policies, and graph operands supply alias information.
The following examples illustrate failures in carrying or
checking these facts.
Each box shows the required fact flow and checks.

\begin{itemize}[leftmargin=1.25em,itemsep=0.6em]

  \item \begin{minipage}[t]{\linewidth}
  \textbf{{IREE: Nonzero-offset buffer view~\cite{tb-iree-14215}.}}
  \begin{rqtwopathbox}[unbreakable,toprule=0.5pt, bottomrule=0.6pt]
    \texttt{buffer view + source offset} $\rightarrow$
    \rqtwocritical{offset-aware fold guard} $\rightarrow$
    \texttt{descriptor range}

    \par\noindent
    The pre-fix fold dropped the nonzero offset.
    The repair guards the fold to preserve the view's
    storage range.
  \end{rqtwopathbox}
  \end{minipage}

  \item \begin{minipage}[t]{\linewidth}
  \textbf{{XLA ecosystem: Copy and donation policy~\cite{tb-jax-24521}.}}
  \begin{rqtwopathbox}[unbreakable,toprule=0.7pt]
    \texttt{may\_alias=False} $\rightarrow$
    \texttt{sharding handler} $\rightarrow$
    \rqtwocritical{C++ transfer policy} $\rightarrow$
    \texttt{destination storage ownership}

    \par\noindent
    The pre-fix sharding path omitted the copy request,
    so the runtime reused its input.
    The repair carries the copy policy into the transfer.
  \end{rqtwopathbox}
  \end{minipage}

  \item \begin{minipage}[t]{\linewidth}
  \textbf{{PyTorch ecosystem:
  Indexed-update reinplacement~\cite{tb-pytorch-197829}.}}
  \begin{rqtwopathbox}[unbreakable]
    {\raggedright
    \texttt{inp, src, \_view\_ops} $\rightarrow$
    \rqtwocritical{\texttt{inp/src} storage-identity test}
    $\rightarrow$
    \texttt{should\_reinplace\_scatter} $\rightarrow$
    \texttt{view(inp).copy\_(src)}\par}

    \par\noindent
    Here, \texttt{inp} is the destination-view base,
    \texttt{src} supplies the values, and
    \texttt{\_view\_ops} constructs the view.
    The pre-fix decision allowed an in-place update without
    checking whether \texttt{inp} and \texttt{src} shared storage.
    The repair compares their storage identities and keeps
    the functional scatter when they match.
  \end{rqtwopathbox}
  \end{minipage}

\end{itemize}

\find{Compiler decisions consume the primary binding fact
in 101 of 149 causes.
The source examples distinguish lost facts from missing checks,
suggesting checks at representation changes and consuming decisions.}

\subsection{RQ3: Binding-State Transitions}\label{sec:rq3}

\begin{figure}[H]
  \centering
  \includegraphics[width=\textwidth]{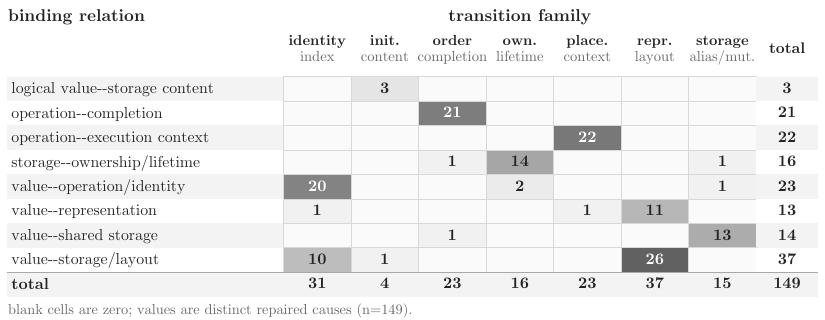}
  \vspace{-15pt}
  \caption{Relation--transition matrix for 149 distinct repaired causes.}
  \Description{A grayscale matrix crosses eight relation families
  with seven transition families. Value--operation/identity
  concentrates in identity or index remapping.
  Value--storage/layout spans representation or layout change,
  identity or index remapping, and initialization or content update.
  Operation--execution context concentrates in placement or context
  change, and value--shared storage in storage, alias, or mutation
  change. Column totals are 31 identity or index remappings,
  four initialization or content updates, 23 ordering or completion
  handoffs, 16 ownership or lifetime changes, 23 placement or context
  changes, 37 representation or layout changes, and 15 storage,
  alias, or mutation changes.}
  \label{fig:rq3-transitions}
\end{figure}

Representation or layout change is the largest transition
family, with 37 of 149 causes (24.8\%), followed by identity
or index remapping with 31 (20.8\%).
Placement or context change and ordering or completion handoff
contribute 23 causes each.
These four families account for 114 causes (76.5\%).
Figure~\ref{fig:rq3-transitions} pairs the transition at each
first invalid decision with its primary binding relation.

The same binding relation can fail under different transitions.
The 37 value--storage/layout causes span representation or
layout change (26), identity or index remapping (10),
and initialization or content update (1).
Conversely, one transition can affect different relations:
identity or index remapping also accounts for 20 of the
23 value--operation/identity causes.
Other relations concentrate in one transition family.
Of the 14 value--shared storage causes, 13 involve storage,
alias, or mutation change.
All 22 operation--execution-context causes involve placement
or context change.
Transition patterns also vary across the sampled ecosystems.
IREE has 11 ownership or lifetime changes involving reference
replacement, allocator lifetime, and asynchronous cleanup.
The PyTorch ecosystem has 23 representation or layout changes
involving view geometry, lowering, and representation selection.
In the XLA ecosystem, 17 causes involve placement or context
changes in device commitment, sharding, streams, collectives,
or callbacks.

A transition can also invalidate a binding while the bound
objects remain present.
In one XLA ecosystem cause, a sharding path dropped the copy
policy while retaining the transfer operation and input buffer.
In a Triton cause, instruction movement reversed the required
order while the tile and barrier remained present.
The failures concern the policy or ordering governing
the objects' use, rather than their presence.

\find{The same binding relation can fail under different
transitions, and one transition can affect several relations.
Their combination guides checks of mappings, policies,
and ordering.}

\subsection{RQ4: Validity Conditions}\label{sec:rq4}

The 149 repaired causes span 15 primary validity-obligation
families (Table~\ref{tab:rq4-obligations}).
Happens-before ordering is the largest, with {21} causes ({14.1\%}),
followed by index and offset preservation (18, 12.1\%)
and identity correspondence (16, 10.7\%).
Together with ownership and lifetime ({12}) and placement
and context consistency (13), these five families account
for 80 causes (53.7\%).
The remaining ten families cover 69 causes.

\begin{table}[t]
  \centering
  \caption{Primary validity obligations in 149 repaired causes.
  Nonzero relation counts are shown.}
  \label{tab:rq4-obligations}
  \footnotesize
  \setlength{\tabcolsep}{3pt}
  \renewcommand{\arraystretch}{.96}
  \begin{tabular}{@{}
    >{\raggedright\arraybackslash}p{.32\linewidth}
    >{\raggedright\arraybackslash}p{.54\linewidth}r@{}}
    \toprule
    Validity obligation & Relation counts & Total \\
    \midrule
    Happens-before ordering
      & operation--completion 17; execution context 1;
        ownership/lifetime {2}; shared storage 1
      & {21} \\
    Index and offset preservation
      & operation/identity 3; representation 1; storage/layout 14
      & 18 \\
    Identity correspondence
      & operation/identity 14; shared storage 1; storage/layout 1
      & 16 \\
    Ownership and lifetime
      & ownership/lifetime {12} & {12} \\
    Placement and context consistency
      & execution context 13 & 13 \\
    Alias independence
      & shared storage 11 & 11 \\
    Layout compatibility
      & storage/layout 11 & 11 \\
    Representation fidelity
      & representation 10; storage/layout 1 & 11 \\
    Range and alignment safety
      & operation/identity 1; storage/layout 9 & 10 \\
    Policy and configuration propagation
      & execution context 5; ownership/lifetime 1;
        shared storage 1 & 7 \\
    Transformation equivalence
      & operation--completion 1; operation/identity 3;
        representation 2; storage/layout 1
      & 7 \\
    Completion and progress
      & operation--completion 3; ownership/lifetime 1
      & 4 \\
    Initialization before publication
      & storage content 3; execution context 1 & 4 \\
    Resource and capability agreement
      & execution context 2; operation/identity 1 & 3 \\
    Value semantics preservation
      & operation/identity 1 & 1 \\
    \bottomrule
  \end{tabular}
\end{table}

Primary obligations vary within binding-relation families.
The 23 value--operation/identity causes span six obligation
families, led by identity correspondence (14),
index and offset preservation (3), and
transformation equivalence (3).
The 37 value--storage/layout causes are concentrated in
index and offset preservation (14), layout compatibility (11),
and range and alignment safety (9).
Other relations show stronger concentrations:
ownership and lifetime is primary in {12} of 16
storage--ownership/lifetime causes, while placement and context
consistency is primary in 13 of 22 execution-context causes.
Transitions further distinguish the conditions required
within the same binding relation.
Among the 37 value--storage/layout causes, eight of ten
identity or index remappings require index and offset
preservation.
The 26 representation or layout changes span five primary
obligations: layout compatibility (11), range and alignment
safety (7), index and offset preservation (6),
identity correspondence (1), and transformation equivalence (1).
The single initialization or content update requires
range and alignment safety.

Source repairs illustrate these differences.
In PyTorch's slice-offset case~\cite{tb-pytorch-108472},
the repair adds the relative offset to the input's
existing storage offset, preserving the intended storage region.
In IREE's host-copy case~\cite{tb-iree-15508},
a two-byte source fits within a four-byte padded target.
The repair replaces a buffer-length equality check with
a check that the source fits within the target.
Both failures affect storage bindings, but their
operations require different validity conditions.

\find{The 149 repaired causes span 15 primary validity
obligations.
Different transitions affecting the same binding relation
require different validity conditions, motivating
transition-specific checks.}

\subsection{RQ5: Consequences and Repair Patterns}\label{sec:rq5}

\subsubsection{Observed Consequences}

Incorrect result or state is the primary consequence in
168 of the 264 reports (Table~\ref{tab:rq5-impact}).
These failures include numerical corruption, stale values,
wrong placement, and incorrect alias or mutation state.

\begin{table}[H]
  \centering
  \caption{Primary consequences in 264 reports.}
  \label{tab:rq5-impact}
  \small
  \renewcommand{\arraystretch}{1.00}
  \begin{tabular}{@{}lr@{}}
    \toprule
    Primary consequence & Reports \\
    \midrule
    Incorrect result or state & 168 \\
    Runtime rejection or crash & 41 \\
    Memory-access violation & 23 \\
    Hang or deadlock & 12 \\
    Resource leak or exhaustion & 12 \\
    Data race & 5 \\
    Other observable failure & 3 \\
    \midrule
    Total & 264 \\
    \bottomrule
  \end{tabular}
\end{table}

Incorrect result or state appears in 33 of 74 IREE reports,
41 of 77 JAX reports, and eight of 14 XLA reports.
The corresponding counts are 65 of 71 PyTorch reports
and 21 of 28 Triton reports.
Repository counts remain separate because the sampling
predicates differ.
PyTorch requires both the Inductor and silent-correctness
labels, while Triton uses its broader bug label.
Sixty reports carry at least one memory-access, concurrent-access,
or availability flag.
These comprise 23 memory-access signals, 14 concurrent-access
violations, and 24 availability losses.
One report carries two flags.
Five concurrent-access violations appear as races;
the others appear as incorrect results or crashes.
These flags capture effects that cross primary
consequence categories.

These patterns motivate output comparisons and explicit state
checks alongside memory, synchronization, and progress checks.
The flags describe technical effects rather than established
security impacts.
Establishing security impact also requires attacker capabilities,
protected assets, and deployment conditions, which the public
reports and linked changes do not specify.

\subsubsection{Failure Families and Repair Strategies}

\begin{figure*}[htp]
  \centering
  \includegraphics[width=\textwidth]{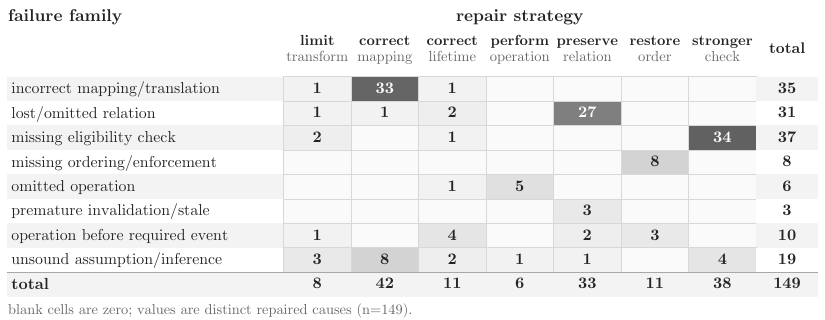}
  \caption{Failure families and repair strategies
  in 149 repaired causes.}
  \Description{A grayscale matrix crosses eight failure families
  with seven repair strategies. Of 35 incorrect mapping or
  translation causes, 33 pair with correct mapping or representation.
  Of 37 missing eligibility check causes, 34 pair with strengthen
  eligibility check. Of 31 lost or omitted relation causes, 27 pair
  with preserve or propagate relation. Repair-strategy totals are:
  constrain unsafe transformation, 8; correct mapping or
  representation, 42; correct ownership or lifetime, 11;
  perform omitted operation, 6; preserve or propagate relation, 33;
  restore ordering or synchronization, 11; and strengthen
  eligibility check, 38.}
  \label{fig:rq5-mechanisms}
\end{figure*}

Three failure families account for 103 of the 149 repaired
causes (69.1\%).
Missing eligibility checks contribute 37 causes (24.8\%),
incorrect mappings or translations contribute 35 (23.5\%),
and lost or omitted relations contribute 31 (20.8\%).
These failures permit operations without checking required
preconditions, establish incorrect correspondences, or fail
to carry required binding facts to their consumers.
Figure~\ref{fig:rq5-mechanisms} pairs each cause's failure
family with its repair strategy.

Of the 103 causes in these families, 94 (91.3\%) use
the corresponding repair categories.
Mapping corrections address 33 of 35 mapping failures,
stronger eligibility checks address 34 of 37 missing-check
failures, and relation preservation or propagation addresses
27 of 31 lost-relation failures.
These correspondences partly reflect the category definitions,
which describe failures and their direct corrections.
The remaining nine causes use different strategies:
four constrain unsafe transformations, four correct ownership
or lifetime, and one corrects a mapping or representation.
Other failure families also span multiple repair strategies.
The \emph{operation before required event} family spans four,
while \emph{unsound assumption or inference} spans six.

\begin{figure}[htp]
  \centering
  \includegraphics[width=\columnwidth]{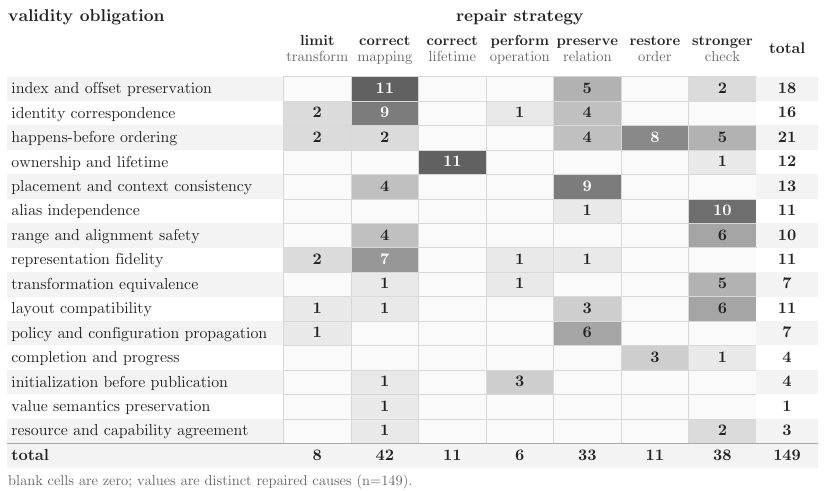}
  \caption{Primary validity obligations and repair strategies
  in 149 repaired causes.}
  \Description{A grayscale matrix crosses 15 obligations with
  seven repair strategies. Mapping correction leads for index
  and offset preservation; restored ordering leads for
  happens-before ordering; ownership correction leads for
  ownership and lifetime; and performing an omitted operation
  leads for initialization before publication.
  Repair-strategy totals are: constrain unsafe transformation, 8;
  correct mapping or representation, 42; correct ownership or
  lifetime, 11; perform omitted operation, 6; preserve or propagate
  relation, 33; restore ordering or synchronization, 11;
  and strengthen eligibility check, 38.}
  \label{fig:rq5-obligation-repair}
\end{figure}

Repair strategies also vary with the violated validity condition
(Figure~\ref{fig:rq5-obligation-repair}).
Mapping correction is the most frequent strategy for index
and offset preservation (11 of 18 causes), while restored
ordering leads for happens-before ordering (8 of 21).
Ownership corrections address 11 of 12 ownership and lifetime
causes, and stronger eligibility checks address ten of
11 alias-independence causes.
For initialization before publication, three of four
repairs add the omitted operation.

Source repairs illustrate why failure mechanisms and repair
strategies need not correspond directly.
In PyTorch's CPU cast case~\cite{tb-pytorch-185337},
\texttt{to\_dtype} caches the original \texttt{float32} value
for reuse after conversion to \texttt{float16}.
A later cast back to \texttt{float32} can therefore bypass
the required rounding.
The repair disables cache insertion under
\texttt{emulate\_precision\_casts}, constraining the optimization
rather than adding an eligibility check.
In IREE's allocator-lifetime case~\cite{tb-iree-9033},
\texttt{iree\_hal\_buffer\_initialize} stores
\texttt{device\_allocator} in the buffer but returns without
retaining it.
The allocator can then be released before buffer destruction,
which still requires it for deallocation.
The first invalid decision is the initializer's return,
and the failure family is \emph{omitted operation}:
the pointer reaches the buffer, but the required retain is absent.
The repair adds the retain and releases the allocator after
deallocation in \texttt{iree\_hal\_buffer\_destroy},
restoring the ownership protocol through its last use.

\find{Three failure families account for 103 of 149 repaired
causes (69.1\%), with corresponding repair categories in
94 cases.
The source examples show that repairs may restore the violated
condition by constraining an unsafe decision or changing
the operations that establish the required binding state.}

\subsubsection{Evidence and Sampling Sensitivity}

All 149 causes have a repair patch and a pre-fix/repaired
source comparison.
Of these, 117 also have a regression test, 89 an attributed
developer diagnosis, and 53 a public reproducer or command.
A cause can have several types of supporting evidence.

Restricting the analysis to the 136 causes with a regression
test or developer diagnosis preserves the leading relation,
transition, obligation, and failure families.
The two leading repair strategies exchange rank:
\emph{strengthen eligibility check} has 38 causes, followed
by \emph{correct mapping or representation} with 37.
Across the 45 relation, transition, obligation, failure,
and repair categories, the largest change in category share
is 2.44 percentage points.
For the 13 causes without a regression test or developer
diagnosis, pre-fix/repaired source comparisons support
their primary obligations.
Six also have a public reproducer.
Repair-linked reports tend to be older and more often closed.
Of the 102 reports without traceable repairs, 67.6\% were
opened in 2025--2026 and 31.4\% were closed.
Among repair-linked reports, the corresponding shares are
51.2\% and 87.7\%.
Incorrect result or state is the leading consequence in
both groups (60.8\% and 65.4\%, respectively).

A further sensitivity analysis excludes the PyTorch repository,
whose query requires the silent-correctness label.
Triton remains included.
Incorrect result or state remains the largest consequence,
with 103 of 193 reports (53.4\%).
Among the remaining 112 repaired causes, the same three
failure families lead, although their order changes:
incorrect mapping or translation (29), lost or omitted
relation ({26}), and missing eligibility check (21).
Together, they account for {76} causes ({67.9\%}).
Compiler decisions consume the primary binding fact in
64 causes (57.1\%), and runtime decisions consume it
in 48 (42.9\%).
Thus, the leading consequence and failure families persist,
and compiler decisions remain the larger consumer group
across the other four repository samples.

\subsection{RQ6: Four Historical Diagnostic Attempts}\label{sec:rq6}

Four archived diagnostic attempts across three IREE causes
illustrate the distinction between reaching a failing path,
detecting and localizing its effect, and explaining its cause.
Each attempt has preserved commands and outputs.
Table~\ref{tab:rq6-diagnostics} summarizes the five dimensions.

\begin{table}[H]
  \centering
  \caption{Diagnostic outcomes for four archived IREE attempts.
  Path: failing-path coverage; Detect: effect detection;
  Effect: effect localization; Cause: cause localization;
  Explain: cause explanation; VMA: Vulkan Memory Allocator;
  N/E: not evaluated.}
  \label{tab:rq6-diagnostics}
  \small
  \renewcommand{\arraystretch}{1.12}
  \setlength{\tabcolsep}{3.2pt}
  \begin{tabular}{@{}lccccc@{}}
    \toprule
    Attempt & Path & Detect & Effect & Cause & Explain \\
    \midrule
    Cached buffer: run-trace~\cite{tb-iree-12161}
      & No  & N/E & N/E & N/E & N/E \\
    Cached buffer: run-module~\cite{tb-iree-12161}
      & No  & N/E & N/E & N/E & N/E \\
    Allocation leak: VMA assertion~\cite{tb-iree-2455}
      & Yes & Yes & Yes & No  & No \\
    Semaphore cleanup: ThreadSanitizer~\cite{tb-iree-9392}
      & Yes & Yes & Yes & Yes & No \\
    \bottomrule
  \end{tabular}
\end{table}

\noindent\textbf{Cached-buffer initialization: path coverage.}
For the cached-buffer failure~\cite{tb-iree-12161},
the archived \texttt{run-trace} attempt compares default
and caching allocators on the same call trace:

\begin{rqcodeblock}[0.78\linewidth]
\begin{lstlisting}[style=rqcode,language=bash]
iree-run-trace --device=local-sync \
  --print_call_outputs calls.yaml > a.txt
iree-run-trace --device=local-sync \
  --print_call_outputs calls.yaml \
  --device_allocator=caching > b.txt
diff a.txt b.txt  # no diff
\end{lstlisting}
\end{rqcodeblock}

The outputs match.
A separate \texttt{run-module} attempt checks one forward call
under default, caching, debug, and caching-plus-debug allocator
configurations.
All four produce the expected outputs.
However, neither attempt executes the repeated-update path
that triggers the failure:

\begin{center}
\begin{rqtwopathbox}[width=0.92\linewidth,breakable=false]
\raggedright
\noindent\textbf{Archived \texttt{run-trace} path.}\enspace
\texttt{calls.yaml input} $\rightarrow$ \texttt{allocate + map}
$\rightarrow$ \texttt{no initial\_data publication}

\par\noindent\textbf{Archived \texttt{run-module} path.}\enspace
\texttt{one forward call} $\rightarrow$
\texttt{input + expected output}
$\rightarrow$ \texttt{no repeated set\_params}

\par\noindent\textbf{Reported failure path.}\enspace
\texttt{set\_params} $\rightarrow$ \texttt{Pack ndarray}
$\rightarrow$ \texttt{pool\_acquire(initial\_data)}
$\rightarrow$ \rqtwocritical{publish reused buffer before write}
\end{rqtwopathbox}
\end{center}

The \texttt{run-trace} replay allocates or maps its inputs,
while \texttt{run-module} performs no repeated parameter updates.
Both bypass the first invalid decision:
\texttt{*out\_buffer = existing\_buffer} publishes reused storage
before writing the supplied \texttt{initial\_data}.
The repair writes those bytes before publication.
Both attempts therefore receive \emph{No} for failing-path
coverage and \emph{not evaluated} for the other four dimensions.

\vspace{3pt}
\noindent\textbf{Allocation leak: effect localization.}
In the allocation-leak case~\cite{tb-iree-2455},
a Vulkan Memory Allocator (VMA) assertion identifies
an outstanding allocation at shutdown.
The allocation is the affected runtime object, and shutdown
is the observation site.
The assertion thus localizes the effect, while subsequent
reference-count tracing and the repair identify the earlier
reference-register clearing operation as the first invalid
decision.

\vspace{3pt}
\noindent\textbf{Semaphore cleanup: effect and cause localization.}
In the semaphore-cleanup case~\cite{tb-iree-9392},
the preserved ThreadSanitizer run reaches the failing path
and reports both conflicting accesses~\cite{clangtsan}:

\begin{rqcodeblock}[0.62\linewidth]
\begin{lstlisting}[style=rqcode]
Read:  task_semaphore.c:392
Write: semaphore_base.c:80
SUMMARY: ThreadSanitizer: data race
\end{lstlisting}
\end{rqcodeblock}

Waiter cleanup reads \texttt{timepoints[i].base.semaphore}
while callback processing clears the embedded field.
The reported read matches the first invalid decision
identified in Section~\ref{sec:method:worked-examples}.
The diagnostic therefore provides both effect and cause
localization, but does not explain the violated condition.
The source contract and repair establish why cleanup
requires a stable identity outside the resettable timepoint.
The repair saves that identity at acquisition and uses it
during cleanup:

\begin{rqcodeblock}[0.78\linewidth]
\begin{lstlisting}[style=rqcode,language=C++]
// After successful timepoint acquisition:
timepoint->semaphore = &semaphore->base;

// During waiter cleanup:
iree_hal_semaphore_t* semaphore = timepoints[i].semaphore;
if (semaphore) {
  iree_hal_semaphore_cancel_timepoint(
      semaphore, &timepoints[i].base);
}
\end{lstlisting}
\end{rqcodeblock}
The violated condition requires ordered access to
callback-mutated state.
By preserving the semaphore identity in a separate field,
the repair eliminates the conflicting read without changing
the callback reset.
Thus, the diagnostic locates the conflicting accesses,
while source evidence explains the invalid decision
and how the repair restores validity.

\find{Two attempts reach the failing path and detect
and localize the physical effect.
One diagnostic output also identifies the selected
invalid-decision site.
Source and repair evidence is required to explain the
violated binding condition.}

\section{Discussion}\label{sec:diss}

\subsection{Implications}

\subsubsection{Identify Transition-Specific Validity Conditions}

Binding relations identify connected entities, while validity
conditions specify what must hold when an operation changes
or consumes their binding.
The joint RQ3--RQ4 analysis identifies distinct requirements
within value--storage/layout: preserving view offsets,
selecting valid reshape strides, and ensuring that copies
fit within target storage.
Representation or layout changes alone span five primary
obligation families within this relation.
The required check therefore depends on the source operation
and its validity condition, not merely the binding relation.

Source reconstructions illustrate these differences.
IREE's cached-buffer reuse requires initialization before
publishing a new value.
PyTorch's scatter reinplacement requires checking whether
\texttt{inp} and \texttt{src} share storage.
IREE's semaphore cleanup requires preserving a stable
identity through callback-driven state changes.
These failures involve different binding facts and operations,
so no single storage, initialization, or ownership check
covers all three.
A binding check therefore needs three elements:
the binding fact, the operation that changes or consumes it,
and the validity condition governing that operation.
Together, they determine what to check and which
implementation state the check requires.
Compiler validators can examine identities, offsets, layouts,
and representations while the relevant IR remains
available~\cite{lopes2021alive2}.
Runtime checks can examine device placement, ownership,
storage, and completion state.
Host and device analyses can provide memory-access and
synchronization information to support these
checks~\cite{clangasan,clangtsan,
jacobson2024hirace,nvidiasanitizer}.

\subsubsection{Preserve Facts and Validate Their Use}

A validity condition cannot be checked if the required
binding fact does not reach its consuming decision.
RQ2 traces one primary fact per cause from its origin
to the operation that changes or consumes it.
The origin and consumer categories differ in 71 causes,
including five whose facts originate in compiler IR or
analysis and are consumed at runtime.
These counts compare provenance endpoints rather than
intermediate layer crossings.

The source examples reveal two failure mechanisms.
A required fact may be lost as its representation changes.
IREE drops a view offset during optimization, while JAX
loses a requested copy policy during an ownership handoff.
Their repairs preserve or propagate the missing facts.
Alternatively, a fact may reach its consumer without
being checked.
PyTorch's scatter reinplacement has access to the operand
storage identities but does not test their aliasing relation
before selecting an in-place update.
These mechanisms require different interventions.
A lost fact must be preserved through the relevant
transformation or handoff, while an available fact can
be checked at its consuming decision.
The source records identify both the representation carrying
the fact and the decision that depends on it,
guiding candidate propagation and checking points.

\subsubsection{Use Failure and Repair Evidence to Guide Checks}

RQ5 connects the condition violated at the first invalid
decision to the operations that restore it.
The condition specifies what must hold, while the repair
identifies how the implementation establishes it.
In IREE's allocator-lifetime case, the invalid decision
occurs when initialization completes without retaining
the allocator.
The repair adds a retain during initialization and a release
after deallocation, preserving ownership through the
allocator's last use.

\begin{samepage}
The observed repairs suggest four forms of candidate checks:

\begin{itemize}[leftmargin=1.25em,itemsep=0.2em]
  \item \textbf{Mapping checks.}
        Verify that remapped positions select the intended values,
        offsets identify the correct storage regions, and layout
        or encoding changes preserve value semantics.

  \item \textbf{Eligibility checks.}
        Verify the preconditions for reuse, fusion, mutation,
        and fast paths.

  \item \textbf{Propagation checks.}
        Verify that required policies, placement, precision,
        identities, and completion state reach their consumers.

  \item \textbf{Ordering and lifetime checks.}
        Verify required completion ordering and resource liveness.
\end{itemize}
\end{samepage}

Observed consequences also inform test-oracle design.
Incorrect result or state is the primary consequence in
168 of 264 reports, motivating output comparisons and
explicit state assertions alongside memory and
synchronization checks.
Evaluating these candidate checks on unseen failures remains
necessary to establish their coverage and precision.

\subsubsection{Measure Diagnostic Outcomes Separately}

RQ6 distinguishes five diagnostic outcomes:
failing-path coverage, effect detection, effect localization,
cause localization, and cause explanation.
Two of the four archived attempts reach the failing path
and detect and localize the physical effect.
One diagnostic output also identifies the selected
invalid-decision site.
Source and repair comparisons establish the violated
binding conditions.
The other two attempts bypass the failing path,
showing why path coverage must be established before
assessing detection.
These distinctions guide diagnostic benchmark design.
The test input and execution trace establish whether
the failing path executes.
Diagnostic output identifies the detected effect and
its reported location.
The issue report, pre-fix source, and repair provide
the reference for assessing cause localization and explanation.
Recording these outcomes separately distinguishes
failure to exercise a path from failure to detect its effect,
and locating an effect from explaining its cause.

\subsection{Threats to Validity}

\subsubsection{Construct validity.}

The taxonomy maps implementation-specific binding relations
and failure mechanisms to shared categories.
Each record retains its source description and normalized label,
and every codebook revision was reapplied to all 149 causes.
The reported distributions use the 2026-10-01 codebook
and depend on its category boundaries.

The artifact preserves joint review decisions for three
pairs of included and excluded reports, but not the
reviewers' separate labels.
The closest scope boundary distinguishes operation-identity
failures from incorrect logical computations.
JAX's \texttt{jnp.maximum} case~\cite{tb-jax-40564}
arises in \texttt{jax.jit(jax.grad(...))}.
The pre-fix \texttt{max\_p} Jacobian--vector product (JVP) rule
multiplies each operand tangent by an \texttt{\_balanced\_eq}
mask comparing the operand with the maximum result.
Reverse-mode differentiation transposes this linearized
computation, so the masks also govern the reported
gradient.\footnote{Pre-fix revision \texttt{f7ef90b3}:
\texttt{jax/\_src/api.py}, lines~551--554, 1712--1714, and 1843--1846;
\texttt{interpreters/ad.py}, lines~342--370, 805--814, and 971--981;
\texttt{lax/lax.py}, lines~5234--5243 and 9833--9846.}
Independently tuned GPU recomputations round differently,
causing both equality masks to become zero and dropping
the winning operand's gradient contribution.
The repair uses \texttt{\_balanced\_cmp} to compare
the operands directly.
We include this case because the incorrect masks lose
the association between the selected operand and its
derivative contribution.
An incorrect contribution remains outside the scope
when its operand association is preserved.

To assess sensitivity to this boundary, removing all 23
value--operation/identity causes leaves 126 causes.
The same three failure families remain largest:
\emph{missing eligibility check} (34),
\emph{lost or omitted relation} (27), and
\emph{incorrect mapping or translation} (25).
They account for 86 of 126 causes (68.3\%).

\subsubsection{Internal validity.}

All 149 repaired causes have a landed repair patch
and a pre-fix/repaired source comparison.
Of these, 136 also have a regression test or an attributed
developer diagnosis.
Two human reviewers checked every cause against its report,
source revisions, tests, and repair.
Their review covered the selected source decision,
physical effect, repair, coded fields, provenance endpoints,
and duplicate links.
Corrections and rechecks produced agreement on all 149
final cause records and the 162-to-149 report mapping.
The archive preserves the final reconciled source operation,
but not each reviewer's initial selection or candidate count.
It therefore supports agreement on the final cause records,
but does not establish an independent localization-agreement rate.
The analysis also follows incorporated repairs rather than
evaluating alternative repair designs.
Failure and repair labels draw on the same source evidence,
and several category pairs describe failures and their
direct corrections.
These definitions partly shape the observed correspondence
in Figure~\ref{fig:rq5-mechanisms}, which summarizes the
landed repairs rather than demonstrating a predictive relation.

\subsubsection{External validity and report selection.}

Repository-specific labels define different sampling frames.
In particular, the PyTorch query requires a silent-correctness
label, while the other repositories use broader defect labels.
The study covers public reports from the selected compiler
and runtime stacks, rather than private defects, other
ecosystems, or a representative sample of deployed workloads.
Human reviewers reversed 28 removal proposals during selection.
The restored reports are already included in the final
264-report analysis set.
A random audit of 600 of the 6,729 initial exclusions
identified four binding failures.
This audit estimates filtering error within the initial
exclusion set.
The calculation uses reported human-review totals, but
the released records do not support independent verification
of sample membership or individual review decisions.
Consequently, the reported distributions characterize
the selected public reports.
Cause-level distributions further depend on the availability
of traceable repairs.

\section{Related Work}\label{sec:related}

Prior work studies how tensor state is represented, how failures
are detected and localized, and how machine-learning software
bugs are classified.
\tech{} focuses on the source decisions and validity conditions
behind tensor binding failures.

\noindent\textbf{Compiler and runtime mechanisms.}
Deep-learning compilers lower programs through graph, tensor,
loop, and target-specific
representations~\cite{li2020dlcompilersurvey,lattner2021mlir}.
XLA, IREE, PyTorch~2, and Triton distribute graph capture,
compilation, buffer planning, kernel generation, and execution
across compiler and runtime components
\cite{openxlaarchitecture,liu2022tinyiree,ansel2024pytorch2,
tillet2019triton}.
These designs determine where binding facts are created,
represented, and consumed.

Several systems preserve or recover relations across these
representations.
COVAN tracks tensor and operator provenance across non-injective
rewrites without requiring every transformation to retain
stable identifiers~\cite{tian2026covan}.
MLIR's One-Shot Bufferize derives alias and equivalence
information from SSA use--def chains.
It uses read-after-write constraints to select in-place
reuse~\cite{mlirbufferization}.
Its ownership-based deallocation pipeline records release
responsibility and can retain runtime alias checks when
static analysis leaves ownership unresolved~\cite{mlirownership}.
\tech{} examines historical failures involving these relations.
For each repaired cause, it connects the binding fact's origin
and consumer to the first invalid decision, validity condition,
physical effect, and landed repair.

\vspace{1pt}
\noindent\textbf{Failure detection and localization.}
Compiler testing combines input generation, execution, oracles,
and failure reduction
\cite{chen2020compilertesting,zhang2025dllibrarytesting}.
Tzer, NNSmith, HirGen, GenCoG, TorchProbe, WhiteFox, and XCheck
generate or mutate programs under different language,
optimization, and platform constraints
\cite{liu2022tzer,liu2023nnsmith,ma2023hirgen,wang2023gencog,
su2023torchprobe,yang2024whitefox,qiu2026xcheck}.
CRADLE compares backend behavior~\cite{pham2019cradle},
while Alive2 and Optimuzz validate compiler
transformations~\cite{lopes2021alive2,kwon2025optimuzz}.
Once a failure is found, DuoReduce minimizes the failing
MLIR program and pass sequence~\cite{wang2025duoreduce}.
Dynamic tools detect memory and synchronization errors
during execution.
AddressSanitizer and ThreadSanitizer cover supported host
memory errors and data races~\cite{clangasan,clangtsan}.
NVIDIA Compute Sanitizer and HiRace target device memory,
initialization, synchronization, and race
errors~\cite{nvidiasanitizer,jacobson2024hirace}.
\tech{} reconstructs the validity condition behind an observed
effect using reports, historical source, tests, and patches.
Its exploratory IREE case study separates failing-path coverage,
effect detection, effect localization, cause localization,
and cause explanation.

\vspace{1pt}
\noindent\textbf{Empirical studies of machine-learning software.}
Framework studies classify symptoms, root causes, affected
components, and repair patterns
\cite{jia2021tensorflow,yang2022frameworkbugs}.
Compiler studies examine TVM, Glow, nGraph, PlaidML,
and Tensor Comprehensions
\cite{du2021compilerbugs,shen2021compilerbugs}.
Orion derives fuzzing rules from historical
vulnerabilities~\cite{harzevili2025orion}, while Huang et al.
study false-positive reports in TVM and
OpenVINO~\cite{huang2025falsepositive}.
Work on indivisible multi-hunk bugs shows that coordinated
edits can be necessary to capture a complete
repair~\cite{xin2024multihunk}.
The closest studies focus on PyTorch.
Li et al. classify 116 closed \texttt{torch.compile}
correctness bugs by component, root cause, and trigger.
They also evaluate testing tools and derive
AlignGuard~\cite{li2026torchcorrectness}.
Their categories include incorrect alias tracking and layout
conflicts.
Yuan et al. study TorchDynamo frontend bugs and targeted
tests~\cite{yuan2026frontendbugs}.
\tech{} covers three ecosystems and includes crashes, leaks,
hangs, and races alongside incorrect results.
Its cause-level unit is a repaired source decision that changes
or consumes a tensor binding fact.
For example, an alias-tracking label captures shared storage
in {PyTorch's scatter-reinplacement case~\cite{tb-pytorch-197829}}.
The \tech{} record identifies the \texttt{inp}/\texttt{src}
storage relation and the decision in
\texttt{should\_reinplace\_scatter} that permits an in-place update.
It links that decision to the alias-independence condition,
overlapping accesses, and storage-identity guard in the repair.

\section{Conclusion}\label{sec:conclude}

Tensor binding failures can originate in one layer and
manifest in another.
\tech{} connects these failures to source decisions,
validity conditions, and repairs across IREE and the
XLA and PyTorch ecosystems.
The study analyzes 264 human-confirmed reports and
149 distinct repaired causes.
Three failure families account for {103} causes ({69.1\%}):
missing eligibility checks, incorrect mappings or translations,
and lost or omitted relations.
The first invalid decision occurs in compiler code for
101 causes and runtime code for 48.
Diagnostic cases further distinguish effect localization
from cause explanation.
These findings motivate checks that preserve binding facts
across representations and enforce validity conditions
at consuming decisions.
The source records guide check design and placement,
while evaluation on unseen failures remains necessary
to establish coverage and precision.

\bibliographystyle{ACM-Reference-Format}
\bibliography{reference}

@STRING{FSE     = "Proceedings of ACM International Symposium on the Foundations of Software Engineering"}

@STRING{PLDI    = "Proceedings of ACM Conference on Programming Language Design and Implementation"}

@misc{cvelistv5,
  title = {{CVE List V5}},
  author = {{CVE Program}},
  howpublished = {\url{https://github.com/CVEProject/cvelistV5}},
  year = {2026}
}

@misc{openxlaarchitecture,
  title = {{XLA} Architecture},
  author = {{OpenXLA Project}},
  year = {2026},
  url = {https://github.com/openxla/xla/blob/main/docs/architecture.md}
}

@misc{jaxdocumentation,
  title = {Writing Custom {Jaxpr} Interpreters in {JAX}},
  author = {{JAX Contributors}},
  year = {2026},
  url = {https://docs.jax.dev/en/latest/notebooks/Writing_custom_interpreters_in_Jax.html}
}

@misc{lattner2021mlir,
  title = {{MLIR}: A Compiler Infrastructure for the End of Moore's Law},
  author = {Lattner, Chris and Amini, Mehdi and Bondhugula, Uday and Cohen, Albert and Davis, Andy and Pienaar, Jacques and Riddle, River and Shpeisman, Tatiana and Vasilache, Nicolas and Zinenko, Oleksandr},
  year = {2020},
  eprint = {2002.11054},
  archivePrefix = {arXiv},
  url = {https://arxiv.org/abs/2002.11054}
}

@misc{liu2022tinyiree,
  title = {{TinyIREE}: An {ML} Execution Environment for Embedded Systems from Compilation to Deployment},
  author = {Liu, Hsin-I Cindy and Brehler, Marius and Ravishankar, Mahesh and Vasilache, Nicolas and Vanik, Ben and Laurenzo, Stella},
  year = {2022},
  eprint = {2205.14479},
  archivePrefix = {arXiv},
  url = {https://arxiv.org/abs/2205.14479}
}

@inproceedings{ansel2024pytorch2,
  title = {{PyTorch 2}: Faster Machine Learning Through Dynamic {Python} Bytecode Transformation and Graph Compilation},
  author = {Ansel, Jason and others},
  booktitle = {Proceedings of the International Conference on Architectural Support for Programming Languages and Operating Systems},
  pages = {929--947},
  publisher = {Association for Computing Machinery},
  address = {New York, NY, USA},
  year = {2024},
  doi = {10.1145/3620665.3640366},
  url = {https://pytorch.org/assets/pytorch2-2.pdf}
}

@inproceedings{tillet2019triton,
  title = {Triton: An Intermediate Language and Compiler for Tiled Neural Network Computations},
  author = {Tillet, Philippe and Kung, H. T. and Cox, David},
  booktitle = {Proceedings of the ACM SIGPLAN International Workshop on Machine Learning and Programming Languages},
  pages = {10--19},
  publisher = {Association for Computing Machinery},
  address = {New York, NY, USA},
  year = {2019},
  url = {https://www.eecs.harvard.edu/~htk/publication/2019-mapl-tillet-kung-cox.pdf}
}

@inproceedings{shen2021compilerbugs,
  title = {A Comprehensive Study of Deep Learning Compiler Bugs},
  author = {Shen, Qingchao and Ma, Haoyang and Chen, Junjie and Tian, Yongqiang and Cheung, Shing-Chi and Chen, Xiang},
  booktitle = {Proceedings of the ACM Joint European Software Engineering Conference and Symposium on the Foundations of Software Engineering},
  pages = {968--980},
  publisher = {Association for Computing Machinery},
  address = {New York, NY, USA},
  year = {2021},
  doi = {10.1145/3468264.3468591}
}

@inproceedings{du2021compilerbugs,
  title = {An Empirical Study on Common Bugs in Deep Learning Compilers},
  author = {Du, Xiaoting and Zheng, Zheng and Ma, Lei and Zhao, Jianjun},
  booktitle = {Proceedings of the IEEE International Symposium on Software Reliability Engineering},
  pages = {184--195},
  publisher = {IEEE},
  address = {Piscataway, NJ, USA},
  year = {2021},
  doi = {10.1109/ISSRE52982.2021.00030}
}

@article{liu2022tzer,
  title = {Coverage-Guided Tensor Compiler Fuzzing with Joint {IR}-Pass Mutation},
  author = {Liu, Jiawei and Wei, Yuxiang and Yang, Sen and Deng, Yinlin and Zhang, Lingming},
  journal = {Proceedings of the ACM on Programming Languages},
  volume = {6},
  number = {OOPSLA1},
  pages = {1--26},
  doi = {10.1145/3527317},
  year = {2022},
  url = {https://arxiv.org/abs/2202.09947}
}

@inproceedings{pham2019cradle,
  title = {{CRADLE}: Cross-Backend Validation to Detect and Localize Bugs in Deep Learning Libraries},
  author = {Pham, Hung Viet and Lutellier, Thibaud and Qi, Weizhen and Tan, Lin},
  booktitle = {Proceedings of the International Conference on Software Engineering},
  pages = {1027--1038},
  publisher = {IEEE},
  address = {Piscataway, NJ, USA},
  year = {2019},
  doi = {10.1109/ICSE.2019.00107}
}

@misc{qiu2026xcheck,
  title = {Finding Compiler-Platform Interaction Bugs in Deep Learning Pipelines via Cross-Layer Constraints},
  author = {Qiu, Yuxin and Wang, Jiyuan and Badhe, Ronak and Limpanukorn, Ben and Kim, Miryung and Zhang, Qian},
  year = {2026},
  eprint = {2606.18421},
  archivePrefix = {arXiv},
  url = {https://arxiv.org/abs/2606.18421}
}

@misc{yuan2026frontendbugs,
  title = {Demystifying Deep Learning Compiler Frontend Bugs: An {LLM}-Aided Empirical Study},
  author = {Yuan, Xinyi and Chen, Wei and Liu, Jinyi and Chen, Pengyu and Wei, Jun and Wu, Guoquan and Zhu, Jiaxin and Huang, Tao},
  year = {2026},
  eprint = {2607.25651},
  archivePrefix = {arXiv},
  url = {https://arxiv.org/abs/2607.25651}
}

@misc{li2026torchcorrectness,
  title = {Demystifying the Silence of Correctness Bugs in {PyTorch} Compiler},
  author = {Li, Meiziniu and Li, Dongze and Liu, Jianmeng and Cheung, Shing-Chi},
  year = {2026},
  eprint = {2604.08720},
  archivePrefix = {arXiv},
  doi = {10.48550/arXiv.2604.08720},
  url = {https://arxiv.org/abs/2604.08720}
}

@misc{clangasan,
  title = {{AddressSanitizer}},
  author = {{LLVM Project}},
  year = {2026},
  url = {https://clang.llvm.org/docs/AddressSanitizer.html}
}

@misc{clangtsan,
  title = {{ThreadSanitizer}},
  author = {{LLVM Project}},
  year = {2026},
  url = {https://clang.llvm.org/docs/ThreadSanitizer.html}
}

@misc{nvidiasanitizer,
  title = {{Compute Sanitizer}},
  author = {{NVIDIA}},
  year = {2026},
  url = {https://docs.nvidia.com/compute-sanitizer/ComputeSanitizer/index.html}
}

@article{li2020dlcompilersurvey,
  title = {The Deep Learning Compiler: A Comprehensive Survey},
  author = {Li, Mingzhen and Liu, Yi and Liu, Xiaoyan and Sun, Qingxiao and You, Xin and Yang, Hailong and Luan, Zhongzhi and Gan, Lin and Yang, Guangwen and Qian, Depei},
  journal = {IEEE Transactions on Parallel and Distributed Systems},
  volume = {32},
  number = {3},
  pages = {708--727},
  year = {2021},
  doi = {10.1109/TPDS.2020.3030548}
}

@article{chen2020compilertesting,
  title = {A Survey of Compiler Testing},
  author = {Chen, Junjie and Patra, Jibesh and Pradel, Michael and Xiong, Yingfei and Zhang, Hongyu and Hao, Dan and Zhang, Lu},
  journal = {ACM Computing Surveys},
  volume = {53},
  number = {1},
  articleno = {4},
  numpages = {36},
  year = {2020},
  doi = {10.1145/3363562}
}

@article{zhang2025dllibrarytesting,
  title = {Deep Learning Library Testing: Definition, Methods and Challenges},
  author = {Zhang, Xiaoyu and Jiang, Weipeng and Shen, Chao and Li, Qi and Wang, Qian and Lin, Chenhao and Guan, Xiaohong},
  journal = {ACM Computing Surveys},
  volume = {57},
  number = {7},
  articleno = {187},
  numpages = {37},
  year = {2025},
  doi = {10.1145/3716497}
}

@inproceedings{liu2023nnsmith,
  title = {{NNSmith}: Generating Diverse and Valid Test Cases for Deep Learning Compilers},
  author = {Liu, Jiawei and Lin, Jinkun and Ruffy, Fabian and Tan, Cheng and Li, Jinyang and Panda, Aurojit and Zhang, Lingming},
  booktitle = {Proceedings of the International Conference on Architectural Support for Programming Languages and Operating Systems},
  pages = {530--543},
  publisher = {Association for Computing Machinery},
  address = {New York, NY, USA},
  year = {2023},
  doi = {10.1145/3575693.3575707}
}

@inproceedings{ma2023hirgen,
  title = {Fuzzing Deep Learning Compilers with {HirGen}},
  author = {Ma, Haoyang and Shen, Qingchao and Tian, Yongqiang and Chen, Junjie and Cheung, Shing-Chi},
  booktitle = {Proceedings of the ACM SIGSOFT International Symposium on Software Testing and Analysis},
  pages = {248--260},
  publisher = {Association for Computing Machinery},
  address = {New York, NY, USA},
  year = {2023},
  doi = {10.1145/3597926.3598053}
}

@inproceedings{wang2023gencog,
  title = {{GenCoG}: A {DSL}-Based Approach to Generating Computation Graphs for {TVM} Testing},
  author = {Wang, Zihan and Nie, Pengbo and Miao, Xinyuan and Chen, Yuting and Wan, Chengcheng and Bu, Lei and Zhao, Jianjun},
  booktitle = {Proceedings of the ACM SIGSOFT International Symposium on Software Testing and Analysis},
  pages = {904--916},
  publisher = {Association for Computing Machinery},
  address = {New York, NY, USA},
  year = {2023},
  doi = {10.1145/3597926.3598105}
}

@inproceedings{su2023torchprobe,
  title = {{TorchProbe}: Fuzzing Dynamic Deep Learning Compilers},
  author = {Su, Qidong and Geng, Chuqin and Pekhimenko, Gennady and Si, Xujie},
  booktitle = {Proceedings of the Asian Symposium on Programming Languages and Systems},
  pages = {310--331},
  publisher = {Springer},
  address = {Cham, Switzerland},
  year = {2023},
  doi = {10.1007/978-981-99-8311-7_15}
}

@article{yang2024whitefox,
  title = {{WhiteFox}: White-Box Compiler Fuzzing Empowered by Large Language Models},
  author = {Yang, Chenyuan and Deng, Yinlin and Lu, Runyu and Yao, Jiayi and Liu, Jiawei and Jabbarvand, Reyhaneh and Zhang, Lingming},
  journal = {Proceedings of the ACM on Programming Languages},
  volume = {8},
  number = {OOPSLA2},
  articleno = {296},
  numpages = {27},
  year = {2024},
  doi = {10.1145/3689736}
}

@inproceedings{lopes2021alive2,
  title = {{Alive2}: Bounded Translation Validation for {LLVM}},
  author = {Lopes, Nuno P. and Lee, Juneyoung and Hur, Chung-Kil and Liu, Zhengyang and Regehr, John},
  booktitle = {Proceedings of the ACM SIGPLAN International Conference on Programming Language Design and Implementation},
  pages = {65--79},
  publisher = {Association for Computing Machinery},
  address = {New York, NY, USA},
  year = {2021},
  doi = {10.1145/3453483.3454030}
}

@inproceedings{jacobson2024hirace,
  title = {{HiRace}: Accurate and Fast Data Race Checking for {GPU} Programs},
  author = {Jacobson, John and Burtscher, Martin and Gopalakrishnan, Ganesh},
  booktitle = {Proceedings of the International Conference for High Performance Computing, Networking, Storage and Analysis},
  articleno = {36},
  numpages = {14},
  publisher = {IEEE},
  address = {Piscataway, NJ, USA},
  year = {2024},
  doi = {10.1109/SC41406.2024.00042}
}

@article{jia2021tensorflow,
  title = {The Symptoms, Causes, and Repairs of Bugs Inside a Deep Learning Library},
  author = {Jia, Li and Zhong, Hao and Wang, Xiaoyin and Huang, Linpeng and Lu, Xuansheng},
  journal = {Journal of Systems and Software},
  volume = {177},
  pages = {110935},
  year = {2021},
  doi = {10.1016/j.jss.2021.110935}
}

@article{harzevili2025orion,
  title = {History-Driven Fuzzing for Deep Learning Libraries},
  author = {Shiri Harzevili, Nima and Mohajer, Mohammad Mahdi and Wei, Moshi and Pham, Hung Viet and Wang, Song},
  journal = {ACM Transactions on Software Engineering and Methodology},
  volume = {34},
  number = {1},
  articleno = {19},
  numpages = {29},
  year = {2025},
  doi = {10.1145/3688838}
}

@article{yang2022frameworkbugs,
  title = {A Comprehensive Empirical Study on Bug Characteristics of Deep Learning Frameworks},
  author = {Yang, Yilin and He, Tianxing and Xia, Zhilong and Feng, Yang},
  journal = {Information and Software Technology},
  volume = {151},
  pages = {107004},
  year = {2022},
  doi = {10.1016/j.infsof.2022.107004}
}

@article{baltes2022sampling,
  title = {Sampling in Software Engineering Research: A Critical Review and Guidelines},
  author = {Baltes, Sebastian and Ralph, Paul},
  journal = {Empirical Software Engineering},
  volume = {27},
  number = {4},
  articleno = {94},
  numpages = {31},
  year = {2022},
  doi = {10.1007/s10664-021-10072-8}
}

@book{cochran1977sampling,
  title = {Sampling Techniques},
  author = {Cochran, William G.},
  edition = {3},
  publisher = {John Wiley \& Sons},
  address = {New York, NY, USA},
  year = {1977},
  isbn = {9780471162407}
}

@article{wang2015hypergeometric,
  title = {Exact Optimal Confidence Intervals for Hypergeometric Parameters},
  author = {Wang, Weizhen},
  journal = {Journal of the American Statistical Association},
  volume = {110},
  number = {512},
  pages = {1491--1499},
  year = {2015},
  doi = {10.1080/01621459.2014.966191}
}

@article{cohen1960coefficient,
  title = {A Coefficient of Agreement for Nominal Scales},
  author = {Cohen, Jacob},
  journal = {Educational and Psychological Measurement},
  volume = {20},
  number = {1},
  pages = {37--46},
  year = {1960},
  doi = {10.1177/001316446002000104}
}

@book{manning2008ir,
  title = {Introduction to Information Retrieval},
  author = {Manning, Christopher D. and Raghavan, Prabhakar and Sch{\"u}tze, Hinrich},
  publisher = {Cambridge University Press},
  address = {Cambridge, UK},
  year = {2008},
  doi = {10.1017/CBO9780511809071},
  isbn = {9780521865715}
}

@inproceedings{huotala2024llmscreening,
  title = {The Promise and Challenges of Using {LLM}s to Accelerate the Screening Process of Systematic Reviews},
  author = {Huotala, Aleksi and Kuutila, Miikka and Ralph, Paul and M\"antyl\"a, Mika},
  booktitle = {Proceedings of the 28th International Conference on Evaluation and Assessment in Software Engineering},
  pages = {262--271},
  publisher = {ACM},
  address = {New York, NY, USA},
  year = {2024},
  doi = {10.1145/3661167.3661172}
}

@misc{tian2026covan,
  title = {Provenance Tracking in {AI} Compilers through the Lens of Coalgebra},
  author = {Tian, Zilu and Liu, Liying},
  howpublished = {arXiv:2606.10937},
  year = {2026},
  doi = {10.48550/arXiv.2606.10937}
}

@misc{mlirbufferization,
  title = {Bufferization},
  author = {{LLVM Project}},
  year = {2026},
  howpublished = {MLIR Documentation},
  url = {https://mlir.llvm.org/docs/Bufferization/}
}

@misc{mlirownership,
  title = {Ownership-Based Buffer Deallocation},
  author = {{LLVM Project}},
  year = {2026},
  howpublished = {MLIR Documentation},
  url = {https://mlir.llvm.org/docs/OwnershipBasedBufferDeallocation/}
}

@article{kwon2025optimuzz,
  title = {Optimization-Directed Compiler Fuzzing for Continuous Translation Validation},
  author = {Kwon, Jaeseong and Jang, Bongjun and Lee, Juneyoung and Heo, Kihong},
  journal = {Proceedings of the ACM on Programming Languages},
  volume = {9},
  number = {PLDI},
  articleno = {172},
  numpages = {24},
  year = {2025},
  doi = {10.1145/3729275}
}

@article{wang2025duoreduce,
  title = {{DuoReduce}: Bug Isolation for Multi-Layer Extensible Compilation},
  author = {Wang, Jiyuan and Qiu, Yuxin and Limpanukorn, Ben and Kang, Hong Jin and Zhang, Qian and Kim, Miryung},
  journal = {Proceedings of the ACM on Software Engineering},
  volume = {2},
  number = {FSE},
  pages = {647--667},
  publisher = {Association for Computing Machinery},
  year = {2025},
  doi = {10.1145/3715747}
}

@article{huang2025falsepositive,
  title = {False-Positive Bug Reports in Deep Learning Compilers: Stages, Root Causes, and Mitigation},
  author = {Huang, Lili and Shen, Qingchao and Wang, Dong and Wu, Yunping and Wang, Meng and Chen, Junjie},
  journal = {ACM Transactions on Software Engineering and Methodology},
  volume = {35},
  number = {8},
  pages = {1--32},
  year = {2026},
  doi = {10.1145/3774889}
}

@article{xin2024multihunk,
  title = {Detecting, Creating, Repairing, and Understanding Indivisible Multi-Hunk Bugs},
  author = {Xin, Qi and Wu, Haojun and Tang, Jinran and Liu, Xinyu and Reiss, Steven P. and Xuan, Jifeng},
  journal = {Proceedings of the ACM on Software Engineering},
  volume = {1},
  number = {FSE},
  pages = {2747--2770},
  numpages = {24},
  year = {2024},
  doi = {10.1145/3660828}
}

@misc{github-search-api,
  author = {{GitHub}},
  title = {{REST API Endpoints for Search}},
  year = {2026},
  howpublished = {GitHub Docs},
  url = {https://docs.github.com/en/rest/search/search?apiVersion=2022-11-28}
}

@misc{tb-iree-12161,
  author = {{sogartar}},
  title = {Caching allocator causes wrong computation result},
  year = {2023},
  howpublished = {IREE issue \#12161. Archived report and repair},
  url = {https://github.com/awen-li/TensorBound/tree/37af7ddfe865f52333fef30d17d20b7e8c927d77/data/final/iree/12161}
}

@misc{tb-iree-9392,
  author = {{bjacob}},
  title = {Data race in {\texttt{iree\_hal\_task\_semaphore\_multi\_wait}}},
  year = {2022},
  howpublished = {IREE issue \#9392. Archived report and repair},
  url = {https://github.com/awen-li/TensorBound/tree/b1ba851239c343799d8678f29465fe5f90ec4b1c/data/final/iree/9392}
}

@misc{tb-iree-2455,
  author = {{antiagainst}},
  title = {Leaked buffers for unused inputs},
  year = {2020},
  howpublished = {IREE issue \#2455. Archived report and repair},
  url = {https://github.com/awen-li/TensorBound/tree/37af7ddfe865f52333fef30d17d20b7e8c927d77/data/final/iree/2455}
}

@misc{tb-iree-14215,
  author = {{ezhulenev}},
  title = {{[Stream]} Buffer views produce stream resources with incorrect offset and size},
  year = {2023},
  howpublished = {IREE issue \#14215. Archived report and repair},
  url = {https://github.com/awen-li/TensorBound/tree/37af7ddfe865f52333fef30d17d20b7e8c927d77/data/final/iree/14215}
}

@misc{tb-iree-15508,
  author = {{qedawkins}},
  title = {{[Python]} Failure to send a tensor of {$<$} 4 bytes to host},
  year = {2023},
  howpublished = {IREE issue \#15508. Archived report and repair},
  url = {https://github.com/awen-li/TensorBound/tree/37af7ddfe865f52333fef30d17d20b7e8c927d77/data/final/iree/15508}
}

@misc{tb-iree-9033,
  author = {{makslevental}},
  title = {{\texttt{iree\_hal\_allocator\_deallocate\_buffer}} segfault with minimal script},
  year = {2022},
  howpublished = {IREE issue \#9033. Archived report and repair},
  url = {https://github.com/awen-li/TensorBound/tree/b1ba851239c343799d8678f29465fe5f90ec4b1c/data/final/iree/9033}
}

@misc{tb-pytorch-108472,
  author = {{jansel}},
  title = {Silent correctness issue if input tensors have existing storage offsets},
  year = {2023},
  howpublished = {PyTorch issue \#108472. Archived report and repair},
  url = {https://github.com/awen-li/TensorBound/tree/37af7ddfe865f52333fef30d17d20b7e8c927d77/data/final/pytorch/108472}
}

@misc{tb-pytorch-197829,
  author = {{alepot55}},
  title = {{[inductor]} {\texttt{x[1:] = x[:-1].clone()}} compiles to a copy over overlapping memory: silent wrong results on {CUDA} and {CPU}},
  year = {2026},
  howpublished = {PyTorch issue \#197829. Archived report and repair},
  url = {https://github.com/awen-li/TensorBound/tree/37af7ddfe865f52333fef30d17d20b7e8c927d77/data/final/pytorch/197829}
}

@misc{tb-pytorch-185337,
  author = {{rookieLiu2018}},
  title = {{[inductor]} {CPU} codegen does not preserve explicit float32 to float16 to float32 cast semantics},
  year = {2026},
  howpublished = {PyTorch issue \#185337. Archived report and repair},
  url = {https://github.com/awen-li/TensorBound/tree/37af7ddfe865f52333fef30d17d20b7e8c927d77/data/final/pytorch/185337}
}

@misc{tb-jax-24521,
  author = {{ojw28}},
  title = {{INVALID\_ARGUMENT} when copying array in {\texttt{pinned\_host}} memory with {\texttt{device\_put(..., may\_alias=False)}}},
  year = {2024},
  howpublished = {JAX issue \#24521. Archived report and repairs},
  url = {https://github.com/awen-li/TensorBound/tree/37af7ddfe865f52333fef30d17d20b7e8c927d77/data/final/jax/24521}
}

@misc{tb-jax-40564,
  author = {{timoklein}},
  title = {{\texttt{jnp.maximum}} {JVP} silently drops a gradient branch on {GPU} when the fusion autotuner emits the operand twice with different rounding},
  year = {2026},
  howpublished = {JAX issue \#40564. Archived report and repair},
  url = {https://github.com/awen-li/TensorBound/tree/37af7ddfe865f52333fef30d17d20b7e8c927d77/data/final/jax/40564}
}

\appendixcontent

\appendix
\raggedbottom

\section[Sampling-Frame Specification]{Sampling-Frame Specification}
\label{app:collection-frame}

The GitHub collector uses the Search Issues endpoint with
\texttt{sort=created}, \texttt{order=asc}, and 100 records
per page.
The whole-range query is:
\begin{center}
\small
\texttt{repo:\{repository\} is:issue}\\
\texttt{created:<=2026-09-24 label:"\{label-1\}"}\\
\texttt{[label:"\{label-2\}"]}
\end{center}
Repository and label values follow
Table~\ref{tab:collection-frame}.
GitHub's Search API returns at most 1,000 results per
query~\cite{github-search-api}.
The IREE and JAX queries reported 1,869 and 3,908 matches,
respectively.
To retrieve all matching issues, the collector divides these
queries into disjoint creation-year intervals, from
\texttt{2008-01-01..2008-12-31} through
\texttt{2026-01-01..2026-09-24}, and retrieves every page
within each interval.
The largest intervals contain 386 IREE and 728 JAX issues,
both below the API limit.
For each repository, the retrieved count equals the sum
of its interval totals.
The collection log also retains empty intervals.
The other three repositories use the whole-range query.
After deduplication, IREE, PyTorch, Triton, JAX, and XLA
contribute 1,869, 248, 510, 3,908, and 281 issues,
respectively, totaling 6,816.

The CVE collector searches GitHub's code index for matching
paths under \texttt{cves/}.
It retrieves each returned record from the official CVE List
at revision
\texttt{308dee2f7b}%
\allowbreak\texttt{a40b22f867}%
\allowbreak\texttt{2f752077a6}%
\allowbreak\texttt{a9b755f9c1}.
Each search term uses:
\begin{center}
\small
\texttt{repo:CVEProject/cvelistV5 path:cves}\\
\texttt{"\{term\}"}
\end{center}

\begin{table}[H]
  \centering
  \small
  \caption{CVE List search results by query term.}
  \vspace{-10pt}
  \label{tab:cve-query-yields}
  \begin{tabular}{@{}lrr@{}}
    \toprule
    Term & Reported matches & Retrieved paths \\
    \midrule
    \texttt{IREE}    & 75  & 75  \\
    \texttt{PJRT}    & 0   & 0   \\
    \texttt{PyTorch} & 194 & 181 \\
    \texttt{Triton}  & 111 & 110 \\
    \texttt{XLA}     & 332 & 326 \\
    \texttt{jaxlib}  & 1   & 1   \\
    \midrule
    Total           & 713 & 693 \\
    \bottomrule
  \end{tabular}
\end{table}

The collector retrieves all returned search pages.
Path discovery uses the live index at collection time,
while the fixed revision determines each record's contents.
The searches reported 713 term-level matches and returned
693 paths (Table~\ref{tab:cve-query-yields}).
The 20 unreturned matches comprise 13 for \texttt{PyTorch},
one for \texttt{Triton}, and six for \texttt{XLA}.
The released CVE inventory contains 693 distinct source
identifiers and paths, with no overlap among the retrieved
records.
Their capture timestamps range from 02:45:33 to 03:15:45 UTC
on September 25, 2026.
These timestamps record retrieval time, separately from the
fixed CVE revision and GitHub issue-creation cutoff.

\section[Sampling-Frame Screening Prompt]{Sampling-Frame Screening Prompt}
\label{app:screening-prompt}

The fixed version-1.0 prompt classifies each record as
\emph{include}, \emph{uncertain}, or \emph{exclude}.
Inclusion requires cited evidence of a logical-value--runtime
relation, its violation, and an observed consequence.
An uncertain record describes a concrete failure, identifies
an in-scope storage or execution entity, and proposes a
connection, but leaves one causal link unresolved.
A symptom alone is excluded unless the record identifies
a source operation that changes or consumes a binding relation.

The boundary rules distinguish binding failures from arithmetic,
indexing, shape, unsupported-operation, compiler-bookkeeping,
generic-allocation, performance-only, and non-defect records.
Each decision cites exact spans from the title, labels,
or report text.
The JSON output records the decision, exclusion code,
relation, violation, consequence, cited spans, missing link,
rationale, and requested follow-up evidence.
Appendix~\ref{app:prompt-execution} describes the output
validation checks.

\section[Final Review Prompt]{Final Review Prompt}
\label{app:validation-prompt}

The final review applies the same scope and evidence rules
using a binary decision.
An in-scope decision cites the binding relation, violation,
and consequence.
An out-of-scope decision cites the exclusion evidence
and identifies which element of the binding-failure definition
is not established.
The prompt covers boundary cases involving pre-execution
rejection, unsupported configurations, generic allocation
pressure, arithmetic or type errors, symptom-only reports,
and user contract violations.

\section{Prompt Execution Record}
\label{app:prompt-execution}

Table~\ref{tab:review-roles} summarizes the review stages,
decision sources, and human coverage.
The archived model runs used OpenAI ChatGPT Codex under
the alias \texttt{gpt-daybreak-blue-latest} for screening,
detailed review, final review, and adjudication.
All recorded runs used high reasoning effort.
The two full-frame screening runs used Codex CLI 0.155.0
on September 26--27, 2026.
Both used the same hashed prompt, codebook, and output schema,
with peer outputs hidden.
For Phase~1 validation, two human reviewers independently
applied the final codebook.
Their evidence packets omitted model labels and rationales,
selection status, and the other reviewer's decision.
Both reviewers later checked all 149 cause records against
their reports, source revisions, tests, and repairs.
After corrections and rechecks, both confirmed every final
cause record.

\begin{table}[t]
  \centering
  \caption{Review stages, evidence, and human coverage.}
  \label{tab:review-roles}
  \footnotesize
  \renewcommand{\arraystretch}{.98}
  \setlength{\tabcolsep}{3pt}
  \begin{tabular}{@{}
    >{\raggedright\arraybackslash}p{.13\linewidth}
    >{\raggedright\arraybackslash}p{.29\linewidth}
    >{\raggedright\arraybackslash}p{.16\linewidth}
    >{\raggedright\arraybackslash}p{.18\linewidth}
    >{\raggedright\arraybackslash}p{.13\linewidth}@{}}
    \toprule
    Stage & Evidence and blinding & Resolution &
    Human review & Output \\
    \midrule
    Screening
      & Peer output hidden
      & Model adjudication
      & Both review a 600-record sample
      & 780 advanced \\

    Detailed review
      & Peer decisions hidden
      & Model adjudication
      & Both review 366 proposed removals
      & 19 restored; 433 remain \\

    Four rounds and final review
      & Prior labels, selection status, and peer output hidden
      & Model adjudication
      & Both review 178 proposed removals
      & 9 restored; 264 remain \\

    Final-set validation
      & Model labels and peer decisions hidden
      & Discussion of disagreements
      & Both review all 264 reports
      & 264 confirmed \\

    Cause records
      & Reports, source revisions, tests, and repairs
      & Evidence review
      & Both check all 149
      & Confirmed records \\

    Report consequences
      & Report and test evidence
      & Final study coding
      & Cause-record and final report reviews
      & 264 analyzed \\
    \bottomrule
  \end{tabular}
\end{table}

\vspace{3pt}
\noindent\textbf{Replay scope.}
The historical manifests retain run times, retry data,
and hashes for prompts, schemas, inputs, and outputs.
They do not record an immutable model revision, temperature,
top-p value, or seed.
The public release contains final records and review summaries,
while model prompts and intermediate outputs remain outside it.
Consequently, exact replay of the historical model calls
is unavailable.

\vspace{3pt}
\noindent\textbf{Released evidence and numerical checks.}
The artifact contains final cause records, reported
human-review totals, and derived results.
Its result index links RQ tables and matrices to report,
cause, or diagnostic-attempt identifiers.
Scripts recompute the corresponding distributions.
Phase~1 calculations use reported sample and review-step totals,
but sample membership lists, individual human decisions,
and per-step restoration links are unavailable in the release.

A source-location correction log distinguishes pre-fix operations
from repair sites for three JAX causes.
The corrections identify the strong cache used by
\texttt{\_initial\_style\_jaxpr}, the LAPACK function-pointer
call in the Cholesky handler, and the equality mask in
\texttt{\_balanced\_eq}.
The corrected source anchors preserve the taxonomy
labels and RQ counts.
A separate coding-correction log classifies IREE's missing
allocator retain as an omitted operation and identifies
initializer completion as the first invalid decision.
It also assigns happens-before ordering to the semaphore-cleanup
read and relation preservation to its repair.
The reported distributions use these corrected labels,
while the archived A/B labels remain unchanged.
These revision-time corrections are separate from the
original human review.

The analysis scripts use Python's standard library.
Figure generation and manuscript compilation require separate
dependencies documented with the paper sources.
RQ6 uses archived diagnostic commands and outputs.
Release verification checks these records without rerunning
the diagnostics.

\vspace{3pt}
\noindent\textbf{Output validation.}
The validator requires one output per input identifier.
It checks the output schema and 60-word limit,
then matches each cited span byte-for-byte against
the archived input.

\vspace{3pt}
\noindent\textbf{Final review and validation.}
The two model outputs were saved before comparison,
and a separate model review resolved disagreements.
The final model review classified 264 of 269 provisional
records as binding failures.
Both human reviewers independently confirmed all 264 reports,
with full agreement before discussion.
Final-set precision relative to these human labels is
$264/264=100\%$.

\section{Human Review of Excluded Records}
\label{app:selection-audit}

Table~\ref{tab:exclusion-audit-strata} summarizes human review
of proposed removals and sampled initial exclusions.
The initial-filtering audit comprises three non-overlapping
random samples of 200 records.
The other groups contain model-proposed removals before
human review.
Reports identified as binding failures in these groups
return to the candidate set before the next review step.

\begin{table}[H]
  \centering
  \caption{Human review of proposed removals and sampled
  initial exclusions. A report may receive removal proposals
  at multiple steps, so totals may count the same report
  more than once.}
  \label{tab:exclusion-audit-strata}
  \scriptsize
  \renewcommand{\arraystretch}{.96}
  \begin{tabular}{@{}lrrr@{}}
    \toprule
    Group &
    \makecell{Records or\\proposals} &
    \makecell{Reviews\\completed} &
    \makecell{Binding-failure\\decisions} \\
    \midrule
    Later rounds and final review
      & 178 & 178 & 9 \\
    Detailed review: excluded
      & 283 & 283 & 13 \\
    Detailed review: unresolved
      & 83 & 83 & 6 \\
    Initial filtering
      & 6,729 & 600 & 4 \\
    \midrule
    Sum of groups
      & 7,273 & 1,144 & 32 \\
    \bottomrule
  \end{tabular}
\end{table}

Two reviewers independently apply the final codebook
and record their decisions before comparison.
They resolve disagreements through discussion and assign
final labels based on the cited evidence.
Human review reverses 28 of the 544 removal proposals.
The remaining 516 removals leave 264 reports in the
analysis set:
\[
  780-516=264.
\]
The final set already includes the restored reports.
The four binding failures found in the random audit remain
separate from the analysis set.
They contribute to the initial-filtering error estimate
in Section~\ref{sec:method:filtering}.

\section[Taxonomy Codebook]{Taxonomy Codebook}
\label{app:taxonomy-codebook}

The codebook defines the 45 normalized labels used in the RQ tables. Each entry
gives the operational definition applied at the first invalid decision or
repair site.

\begingroup
\footnotesize
\lstset{literate={--}{{\textnormal{\textendash}}}1,
  escapeinside={(*@}{@*)}}
\newcommand{\codebookgroup}[2]{%
  \par\addvspace{.35em}%
  \noindent\textbf{#1} \textit{(#2 families)}\par\nobreak}
\newenvironment{codebookblock}{%
  \par\nobreak\noindent\hspace*{.11\linewidth}%
  \begin{minipage}{.78\linewidth}%
  \lstset{aboveskip=1pt,belowskip=1pt}%
}{%
  \end{minipage}\par%
}

\codebookgroup{Relation}{8}
\begin{codebookblock}
\begin{lstlisting}[style=rqcode]
Logical value--storage content: Bytes and initialization encode a value.
Operation--completion: An operation links to readiness or completion.
Operation--execution context: An operation links to device, stream, or backend.
Storage--ownership/lifetime: Storage links to owners and a valid lifetime.
Value--operation/identity: A value or role maps to an operation or slot.
Value--representation: A value links to type, precision, encoding, or carrier.
Value--shared storage: Storage is shared through aliasing, reuse, or mutation.
Value--storage/layout: Elements map to positions, shapes, strides, or offsets.
\end{lstlisting}
\end{codebookblock}

\codebookgroup{Transition}{7}
\begin{codebookblock}
\begin{lstlisting}[style=rqcode]
Identity or index remapping: Remaps identities, arguments, indices, or slots.
Initialization or content update: Initializes, publishes, or updates content.
Ordering or completion handoff: Changes ordering, readiness, or completion.
Ownership or lifetime change: Transfers ownership or changes liveness.
Placement or context change: Changes device, backend, stream, or context.
Representation or layout change: Changes encoding, type, shape, or stride.
Storage, alias, or mutation change: Changes allocation, sharing, or reuse.
\end{lstlisting}
\end{codebookblock}

\codebookgroup{Validity obligation}{15}
These labels follow the binding-scope test in Section~\ref{sec:study}.
Transformation equivalence and value semantics preservation require an
established relation between a value and its operation, storage,
representation, ownership, or execution. Pure numerical or local
transformation defects fall outside the construct.
\begin{codebookblock}
\begin{lstlisting}[style=rqcode]
(*@Alias independence: Mutation or reuse preserves other observable values.@*)
(*@Completion and progress: Required work progresses to completion.@*)
Happens-before ordering: Conflicting operations follow the required order.
Identity correspondence: The selected entity has the intended role.
Index and offset preservation: Indices and offsets select the intended region.
(*@Initialization before publication: Publishes only initialized content or state.@*)
Layout compatibility: Producer and consumer layout contracts agree.
Ownership and lifetime: A resource remains owned through its last use.
Placement and context consistency: Execution uses the intended context.
Policy and configuration propagation: Required policy reaches each consumer.
Range and alignment safety: Addresses meet bounds and alignment.
Representation fidelity: Conversion preserves the represented meaning.
Resource and capability agreement: Runtime satisfies compiled requirements.
Transformation equivalence: Preserves semantics for valid inputs.
Value semantics preservation: Binding preserves established value behavior.
\end{lstlisting}
\end{codebookblock}

\codebookgroup{Failure}{8}
\begin{codebookblock}
\begin{lstlisting}[style=rqcode]
Incorrect mapping or translation: Maps a fact to a wrong target or encoding.
Lost or omitted relation: A required relation fails to reach its consumer.
(*@Missing eligibility check: Omits a check of a required precondition.@*)
Missing ordering or enforcement: A required ordering edge is absent.
Omitted operation: A required action is absent from the executed path.
Operation before required event: Executes before its prerequisite occurs.
Premature invalidation or stale state: Ends a live relation or leaves it stale.
(*@Unsound assumption or inference: Draws a conclusion the facts cannot support.@*)
\end{lstlisting}
\end{codebookblock}

The following rules distinguish three pairs of closely related
failure families.

\par\smallskip
\noindent\textbf{Missing check versus unsound inference.}
Use \emph{missing eligibility check} when a decision omits
a required predicate check.
Use \emph{unsound assumption or inference} when the evaluated
evidence does not justify the conclusion.
IREE's incomplete multibuffering path omits an eligibility
test. In a separate cause, its allocation-emplacement pass
assumes that result and destination positions still correspond
after skipping a result. The repair stops at the first result
that cannot be emplaced, preserving a contiguous prefix.

\par\smallskip
\noindent\textbf{Lost relation versus omitted operation.}
Use \emph{lost or omitted relation} when a required binding fact
is dropped or not propagated.
Use \emph{omitted operation} when a required action, such as
a write, copy, wait, or retain, is absent.
For example, IREE drops a subspan offset in one cause
and omits a write to a reused buffer in another.
Its allocator-lifetime case~\cite{tb-iree-9033} also omits
an operation: initialization stores the allocator pointer but
completes without the retain needed to keep the allocator alive.

\par\smallskip
\noindent\textbf{Missing ordering versus early operation.}
Use \emph{missing ordering or enforcement} when the path lacks
a required dependency edge.
Use \emph{operation before required event} when an ordering
requirement is represented in the existing state, but an
operation executes before its prerequisite.
IREE's pthread wait lacks an epoch dependency, whereas its
poller retires a task before unlinking it.

\codebookgroup{Repair strategy}{7}
\begin{codebookblock}
\begin{lstlisting}[style=rqcode]
Constrain unsafe transformation: Restrict a transformation to its sound domain.
Correct mapping or representation: Correct the target, position, or encoding.
Correct ownership or lifetime: Preserve ownership through last use.
Perform omitted operation: Add the missing write, copy, conversion, or action.
Preserve or propagate relation: Carry a required fact to its consumer.
Restore ordering or synchronization: Add the required dependency edge.
Strengthen eligibility check: Add or strengthen a required check.
\end{lstlisting}
\end{codebookblock}
\endgroup

\end{document}